# An economic decision-making model of anticipated surprise with dynamic expectation

Ho Ka Chan[1] and Taro Toyoizumi[1,2]

[1] *Laboratory for Neural Computation and Adaptation, RIKEN Center for Brain Science*

[2] *Department of Mathematical Informatics, Graduate School of Information Science and Technology, The University of Tokyo*

Corresponding authors: hoka.chan@riken.jp (HKC); taro.toyoizumi@riken.jp (TT)

## Abstract

When making decisions under risk, people often exhibit behaviors that classical economic theories cannot explain. Newer models that attempt to account for these 'irrational' behaviors often lack neuroscience bases and require the introduction of subjective and problem-specific constructs. Here, we present a decision-making model inspired by the prediction error signals and introspective neuronal replay reported in the brain. In the model, decisions are chosen based on 'anticipated surprise', defined by a nonlinear average of the differences between individual outcomes and a reference point. The reference point is determined by the expected value of the possible outcomes, which can dynamically change during the mental simulation of decision-making problems involving sequential stages. Our model elucidates the contribution of each stage to the appeal of available options in a decision-making problem. This allows us to explain several economic paradoxes and gambling behaviors. Our work could help bridge the gap between decision-making theories in economics and neurosciences.

## Introduction

The expected utility theory (EUT) (Friedman & Savage, 1948) is one of the most well-established models for analyzing decision-making under risk. Nevertheless, throughout the past decades, several experimental works, e.g. Allais (1953), Kahneman & Tversky (1979), have revealed instances where people's choice cannot be explained by EUT. Some of those examples with a higher profile are coined the term 'paradox'. Subsequently, new theories aiming to resolve these paradoxes, e.g. the prospect theory (PT) (Kahneman & Tversky, 1979), the extended cumulative prospect theory (CPT) for actions with multiple outcomes (Tversky & Kahneman, 1992), and the regret theory (RT) (Bleichrodt & Wakker, 2015; Loomes & Sugden, 1982), have been proposed. Despite the effort, the mechanisms behind decision-making which account for behaviors not conforming to EUT is still an open area of research.

A well-known solution to this problem, proposed by PT and CPT, is to adopt a nonlinear probability weighting function. The rationale behind is that overweighting of low probability events could be a plausible explanation for risk-seeking or risk-averse behavior in the face of such events. However, it is often not easy to decouple the effect of subjective probability perception and evaluation of outcomes (Gonzalez & Wu, 1999). Moreover, studies have shown that the form of mapping between objective and subjective probability, or even the very existence of subjective probability perception, could be highly context dependent (Gallistel, Krishan, Miller, & Latham, 2014; Hertwig, Barron, Weber, & Erev, 2004; Wu, Delgado, & Maloney, 2009; Zhang & Maloney, 2012). Hence, creating a model with both probability weighting and evaluation of monetary outcomes could be prone to overfitting to a particular decision-making task.

Another important issue is what the reference point which people use to gauge the attractiveness of outcomes is. While conventional models often use the initial asset position (the baseline level) as the reference point, many newer theories find such a reference point not always satisfactory. For example, in PT and CPT, the reference point is decided on a case-by-case basis, through deciphering the linguistic and contextual nuances of the problem as it was presented to the decision maker. This is mostly done by applying editing rules (Kahneman & Tversky, 1979). However, the application of the these editing rules is subject to interpretation (Birnbaum, 2008). With these issues, even simple decision-making problems can be formulated in many possible ways, which sometimes lead to different predictions of people's behaviors. It is also unclear whether the reference point remains constant as these theories implicitly assume throughout our decision-making process, even if the decision-making problem is static.

Here, we propose an alternative model that explains and predicts people's behavior in decision-making problems. The model does not rely on probability weighting and has a dynamic yet unambiguous reference point. The basis of the model is inspired by some neuroscience findings and ideas from economics. First, people may either consciously or unconsciously anticipate possible outcomes of an action before executing it. It was reported in animals that when planning an action, neurons in the brain exhibit activities that resemble possible task-state sequences (Ólafsdóttir, Bush, & Barry, 2018). Second, such anticipation processes may influence decision-making. Many models in economics explain people's choices by applying different concepts of 'anticipation', e.g. by considering anticipated regret for not choosing other options (Bleichrodt & Wakker, 2015; Loomes & Sugden, 1982); by using the accumulation of anticipated elation or disappointment to explain the preference for immediate or delayed consumption (Loewenstein, 1987); and by separating anticipatory evaluation of outcomes into an objective one and a choice-dependent one, and giving conditions on when this separation may lead to time inconsistency that causes irrational behaviors (Caplin & Leahy, 2001; Koszegi, 2010). Third, dopamine neurons in the brain that encode reward prediction errors, i.e., the deviation of reward outcome from its prediction (or, in layman's term, 'surprise'), are responsible not only for learning but also for motivating actions (Wise, 2004). Combining these ideas, we propose that 'anticipated surprise' due to probabilistic outcomes is an essential component in decision-making. More specifically, we postulate that in decision-making involving multiple options, people pre-emptively compute how much each possible outcome of an option deviates from a reference point corresponding to their expectation, a natural choice of which would be the expected value across all outcomes for that option. This deviation, which we call the anticipated surprise, is then nonlinearly scaled by a 'surprise function' and weighted by the objective probability of the realization of an outcome. In cases where the expected values of all options are the same, the option that maximizes such anticipated surprise is chosen.

When facing a complex decision, it is commonly believed that people, rather than considering all possible outcomes simultaneously, may break down the problem into modules and consider the said modules individually and sequentially (Anderson, 2002; Anderson et al., 2011; Doya, Samejima, Katagiri, & Kawato, 2002). To take this into account, we incorporate sequential branching of anticipation into our model. This may happen during a typical static decision-making problem when some outcomes are much more/less likely than others, or when information about the outcomes are partially revealed (after the decision has been made) in a sequential manner. In contrast to conventional models that work with a fixed reference point, we propose that the above-mentioned reference point is updated from step to step in each sequential branch. In each step, it corresponds to the expected value of the sub-branches of the next steps. The difference between the expected value of the current and the previous step also contributes to the overall anticipated surprise.

In this paper, we first introduce in Section 1 our model in its simplest form without sequential branching and outline the properties of the surprise function which allows the reproduction of the results in Kahneman & Tversky (1979; Problem 3, 4, 7, 8), namely risk-seeking for lotteries and risk-averse otherwise in the gain domain, and the reflection effect. In Section 2, we explain why PT cannot work with general problems involving sequential structure and show how we incorporate sequential branching of anticipation into our model. In Section 3, we show how we can apply sequential branching in different scenarios to explain well-known, seemingly irrational behaviors, e.g. sequential revelation of information, ambiguity aversion, and event splitting, many of which cannot be explained by conventional models like PT. In Section 4, we compare our model to other prominent decision-making models, suggest possible neural mechanisms related to our model, and briefly discuss possible ways to extend our model to consider cases where options have different expected values.

# Section 1: Reproducing the patterns of prospect theory

## 1.1 From expected utility theory to prospect theory

In Kahneman's 1979 paper (Problem 3, 4, 7, 8, and their primed version) in which PT is first proposed, a type of gambling problems, as shown in Table 1, was extensively discussed.

|  | Option 1 | | Option 2 | |
| --- | --- | --- | --- | --- |
|  | Reward | Probability | Reward | Probability |
| Outcome 1 | $\bar{x}/p$ | $p$ | $\bar{x}/p'$ | $p'$ |
| Outcome 2 | 0 | $1-p$ | 0 | $1-p'$ |

Table 1: Outline for the gambling problems in Kahneman's 1979 paper (Problem 3, 4, 7, 8, and their primed version). Without loss of generality, we assumed that $p > p'$.

In these problems, people have to weigh between an option that gives them a comparatively small amount of money with a large probability (Option 1) and one that gives them a larger amount of money with a lower probability (Option 2). In Table 1, the rewards are scaled such that the expected value $\bar{x}$ is the same for both options. In the actual examples in Kahneman's paper, even though this is not always strictly the case, the expected values for both options have always been kept close. We assume that $p > p'$.

Experimentally, it was found that if the reward is positive (i.e. in the gain domain), most people choose Option 1 (i.e. risk aversion) when $p'$ is large and Option 2 (i.e. risk seeking) when $p'$ is small. The preference reverses if the reward is negative (i.e. in the loss domain), which is coined the term 'reflection effect'. This result is inconsistent with the popular EUT (Friedman & Savage, 1948). In EUT, people choose the option that maximizes the expected utility $U = \mathrm{E}(u(x))$, where $u$ is the utility function, typically assumed to be increasing and concave, and $x$ denotes possible outcomes in an option. Kahneman's result revealed that if we pick a large $p'$, Option 1 is chosen, suggesting that $U_1 > U_2$, where $U_a$ is the expected utility for Option $a$. If we now scale the probability for both options with a sufficiently small factor $\epsilon$, Option 2 will instead be chosen in the new problem, suggesting that $U_{1'} < U_{2'}$, where $U_{a'}$ is the expected utility for Option $a$ in the new problem. However, by definition $U_{a'} = \epsilon U_a$. This leads to a contradiction.

One obvious fix to the abovementioned inconsistency is to remove the linear relationship between $U$ and the probability $p$ that leads to $U_{a'} = \epsilon U_a$. This is what PT assumes. In PT, the prospect function $V$ is given by $V = \sum_j \pi\left(p(x_j)\right) v(x_j - \theta)$, where $x_j$ denotes possible outcomes in an option. Note that the outcome $x_j$ is measured with respect to the status quo as the baseline, such that $x_j = 0$ corresponds to the scenario in which there is no absolute gain or loss. It is then compared with a constant reference point $\theta$. Although the baseline reference point of $\theta = 0$ is typically used, a non-zero reference point could be imposed for some problems. The prominent difference with EUT in the gain domain is the application of nonlinearity in the probability $p$ by the introduction of the subjective probability function $\pi$. Specifically, it overweighs small probability ($\pi(p) > p$ for small $p$) and vise versa ($\pi(p) < p$ for large $p$). The function $v$ is given by

$$v(x) = \begin{cases} f(x) \text{ for } x \geq 0 \\ -kf(|x|) \text{ for } x < 0 \end{cases}, \quad (1)$$

using an increasing concave function $f(x)$ defined for $x \geq 0$ and a constant $k > 1$. The qualitative difference between $v$ and $u$ is in the loss domain, where $v$ is a convex function and is amplified by $k$. The convexity of $v$ is to account for the reflection effect described above. The inclusion of the amplifying factor $k$ is to account for the risk-averse behaviors of the decision maker as shown in a well-known experimental observation (Rabin, 2000; Tom,

Fox, Trepel, & Poldrack, 2007) that people resist taking 50-50 gamble for even money (e.g. preferring nothing to having 50% chance of gaining $10 and 50% chance of losing $10).

## 1.2 The anticipated surprise model

We aim to account for the above-mentioned behavioral observations using a simpler model. The model is based on anticipated surprise. We define 'surprise' $z$ by the difference between an individual outcome and the expected value across all outcomes, i.e. $z = x - \mathrm{E}(x)$. Then, the surprise value $\Delta$ of a choice is given by

$$\Delta = \mathrm{E}(\delta(z)), \tag{2}$$

where $\delta$ takes the same form as $v$ in eq. (1) except that an increasing convex function $f$ is used instead. We set $k > 1$ for the risk aversion factor in the definition of $\delta$ as in PT.

The major difference between PT and our model is that we do not make any transformation of $p$. Instead, we use the expected value across all outcomes as the reference point to gauge the value of each outcome. By contrast, PT uses the baseline reference point in most cases except when another constant reference point is imposed. In addition, we transform the surprise by a convex function in the gain domain as opposed to PT, which uses a concave function. The opposite is true for the loss domain.

Our model can reproduce all response patterns observed in the type of problems described in Table 1 for any arbitrary convex function $f$. The details of the proof for a general surprise function $\Delta$ is shown in Appendix 1. In Figure 1, we show how $\Delta$ varies with $p$ using $f(z) = z^{1.5}$ as an example. For the gain domain, $\Delta$ decreases from 0 at $p = 1$ to negative values as $p$ decreases, indicating people's general preference for certainty as opposed to gambles. Nevertheless, as $p$ becomes small, $\Delta$ increases and becomes positive, indicating people's preference for lotteries. When $k = 1$, the switch of preference takes place at $p = 0.5$, which is much larger than what experimental observations in Kahneman's 1979 paper and other works (Somasundaram & Diecidue, 2017) suggest (Figure 1, red solid). By choosing a larger value of $k$, the point of switch can be lowered to a more realistic value of $p$ (Figure 1, red dashed).

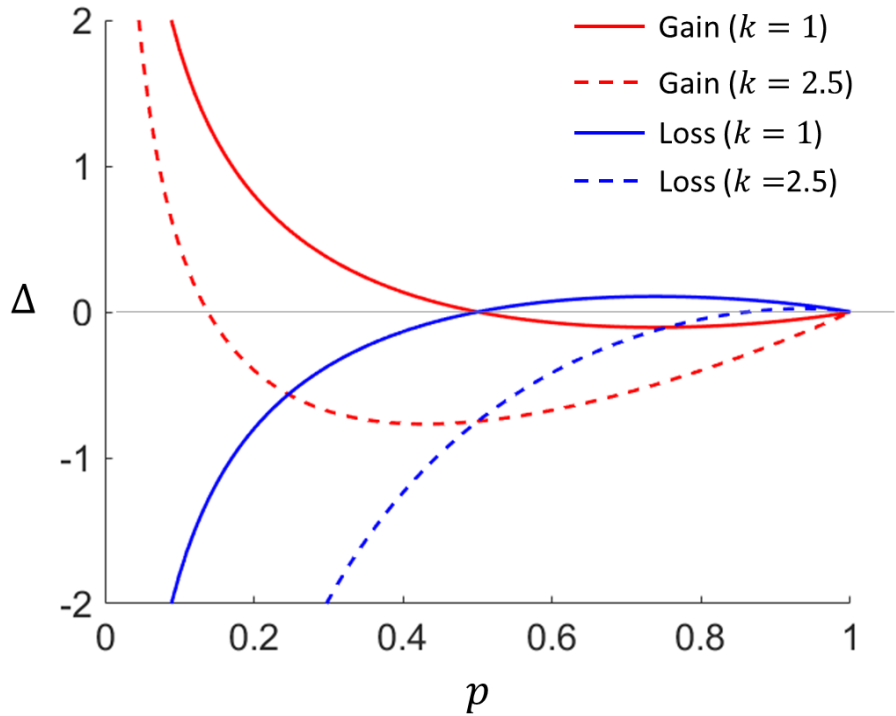

Figure 1: The surprise function $\Delta$ as a function of $p$. In this example, $f(z) = z^{1.5}$. Red (Blue) lines are the results in the gain (loss) domain. For solid lines, $k = 1$. For dashed lines, $k = 2.5$.

In our model, the reflection effect is observed for all values of $p$ when $k = 1$ (See Figure 1, blue solid, and Appendix 1). When $k > 1$, the model can still reproduce the important results of risk aversion (risk seeking) at high $p$ and risk seeking (risk aversion) at low $p$ in the gain domain (in the loss domain, respectively) (Figure 1). However, the reflection effect becomes imperfect, such that the model predicts risk aversion for both the gain and loss domains when $p$ takes intermediate values (around $p = 0.5$ in Figure 1; see Section 4.1 for discussion on the excessive risk aversion in the loss domain when $k > 1$).

Both the proposed model we just presented and PT give similar predictions for the above problem. In the following sections, we study several more general decision-making problems and show that for those problems, the proposed model but not the conventional models (like PT) can account for people's behavior.

## Section 2: Sequential anticipation

### 2.1 Treatment of sequential problem structure by PT

Apart from decision-making with the simple single-stage structure in the previous section, PT also deals with problems in which the process of outcome revelation involves underlying sequential structures (See e.g. Problem 10-12 in Kahneman & Tversky (1979)). Namely, some intermediate outcomes or information about the final outcome are revealed in the midst of generating the final outcome. PT deals with this type of problems by essentially converting them to single-stage problems. The conversion is not unique, and the model prediction can depend on how exactly the sequential structure of the problem is collapsed. For this purpose, PT uses a set of heuristic editing rules that combine identical outcomes, segregate riskless components, or cancel common components across options.  While these editing rules can account for the above examples in Kahneman's paper, they often create ambiguity for general problems because they do not have objective application criteria. In many cases, by applying these rules in different order and combination, the model predicts multiple behavior (Birnbaum, 2008; Kahneman & Tversky, 1979).

Another important feature of PT is that it implicitly assumes a single, static reference point for each problem. This assumption is reasonable if people's expectation does not notably change when they process through different sequential stages mentally. Under this condition, the sequential structure can indeed be effectively collapsed into single-stage as PT does.  Although the baseline reference point works for some problems, different reference points are needed for other problems. In these cases, the reference points are deduced by analyzing the formulation of the options and inferring people's expectations without well-defined procedures, creating ambiguity in the model's predictions.

## 2.2 Intermediate states and update of the reference point

Aside from the issue of ambiguity in PT, people's expectation generally changes as new information about the outcomes is revealed, in contrary to what PT assumes. Imagine your favorite sports team is playing another weaker team. Since your team is stronger, you probably expect your team to win before the match. Now, let us assume that your team is behind in the score midway through the match. At this point, you may already feel disappointed because the probability of your team not winning is higher than it was before the match. If your team indeed loses at the end of the match, you may not be very disappointed at that point because you have already updated your expectation while your team was behind. On the other hand, if your team makes an unlikely comeback win, you may feel pleasantly surprised, again because of the updated expectation.

We aim to formulate our model to closely mimic the thought process as illustrated in the above example. In the example, the score midway through the match constitutes an 'intermediate state'. In these intermediate states, people's expectation on the outcome differs from their initial expectation. We assume that people mentally simulate possible sequences of events comprising the above-mentioned intermediate states during decision-making processes and anticipate the change in expectation upon transition to those intermediate states. Since our model uses the expected outcome as a reference point, the reference point must be updated accordingly. The shift of the reference point would also culminate in intermediate surprise. We will quantify the surprise that emerges from the change in the reference point in the next section. In addition, it is possible that the decision maker does not expect the outcomes to be revealed in steps, but intermediate states are still created during mental simulation simply because of the sequential nature of people's anticipation process or how people group events together. We will discuss these cases with examples in Section 3. As we will show, such intermediate states have a major impact on the anticipated surprise in outcomes that the model computes.

## 2.3 Our model for sequential anticipation

Let us consider the decision-making problem illustrated in Figure 2.

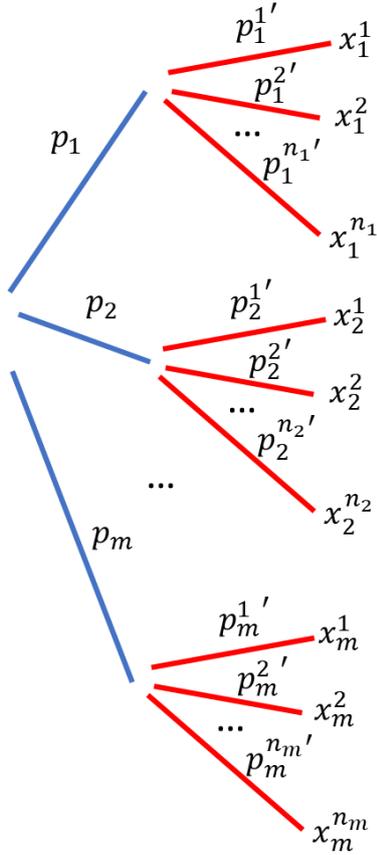

Figure 2: A decision-making problem with sub-branches.

In this two-step problem, there are an intermediate step and an outcome step. In the intermediate step, there are $m$ intermediate states, each happening with a probability $p_i$, $i = 1,2,\ldots m$ as depicted by the blue lines. For each intermediate state, it branches out to $n_i$ possible final outcomes $x_i^j, j = 1,2,3,\ldots,n_i$ with probability $p_i^j$ as depicted by the red lines.

The surprise value ∆ is given by:
$$\Delta = \sum_{i=1}^{m} p_i \delta(E_i - E_0) + \sum_{i=1}^{m} p_i \sum_{j=1}^{n_i} p_i^j \, \delta(x_i^j - E_i), \tag{3}$$

where $E_0 = \sum_i p_i \sum_j p_i^j x_i^j$ is the expected value of all possible outcomes and $E_i = \sum_j p_i^j x_i^j$ is the expected value for the outcomes in the $i^{th}$ sub-branch.

Note that the 1st term in eq. (3) corresponds to the surprise in the intermediate step and the 2nd term corresponds to the surprise in the outcome step. In the model, we assume they contribute additively to the aggregated surprise value.

More generally, $\delta$ can be computed in a cascading manner. Considering a decision-making process with $T$ steps (including the outcome step), we define $y_k(t)$ to be one of the possible state trajectories in our mental branching process, where $t = 0,1,\ldots,T$ refers to the hierarchy of the branch and $k = 1,2,\ldots K$ refers to the index of the trajectory, with $K$ being the total number of trajectories. The surprise value of the $k$th trajectory is given by

$$\Delta_k = \sum_{t=1}^{T} p(y_k(t))\delta(E_k(t) - E_k(t-1)), \tag{4}$$

where $p(y_k(t)) = \prod_{t'=1}^{t} p(y_k(t')|y_k(t'-1))$ is the probability of observing the $k^{\text{th}}$ trajectory up to the $t^{\text{th}}$ step, with $p(y_k(t')|y_k(t'-1))$ referring to the transition probability of entering the state $y_k(t')$ given that the previous state is $y_k(t'-1)$. $E_k(t)$ is the expected value of the outcomes computed at the state $y_k(t)$. Note that $E_k(T)$ is the actual outcome at the end of trajectory $k$.

The aggregated surprise value is then obtained by linear summation of the surprise value of all trajectories

$$\Delta = \sum_{k=1}^{K} \Delta_k \tag{5}$$

Please note that we are not advocating that the branching process would continue infinitely for all decision problems, because of the obvious cognitive burden it presents. How long would the process go on, whether there would be discount for remote branches, and how the sum in eq. (5) can potentially be approximated by subsampling trajectories are beyond the scope of this work.

For some simple problems, the proposed model and PT give the same predictions. For example, the results of problems 10-12 of Kahneman's 1979 paper are also reproduced by the proposed model even though we do not assume people's expectation stays unchanged at the intermediate state, because adding a common first stage to the problem of Table 1 does not change the choice according to eq. (3) (See Appendix 2 for the analysis of our model predictions for problem 10 as an example). However, the model predictions are different with and without the collapsing of sequential branches for general problems. In the following sections, we will present empirical observations that conventional models cannot reproduce and show how our proposed model can account for them. We will specifically describe 3 types of situations where sequential branching makes sense and can explain behaviors that are otherwise hard to understand. They are 1. partial and sequential revelation of outcomes; 2. ambiguity; and 3. restructuring of the problem into probable and improbable outcomes.

# Section 3: Applications of the sequential branching model

## 3.1 Partial revelation of outcomes: blackjack gambling

In blackjack gambling, there are two phases of play: the player phase and the dealer phase. At the start of the player phase, the player makes their decision based on the available information on their and the dealer's hand. At the end of the player phase, the player may receive new information on their hand depending on their earlier decision. While the final outcome may still be uncertain, the probability of the player winning may have changed, which would lead to an update of the expectation of the player. We will illustrate how our sequence branching model captures the effect of such an intermediate state and leads to correct prediction on some observed gambling behaviors. Please refer to Appendix 3 for the relevant details of the blackjack rules.

Bennis (2004) outlined 3 scenarios in which experienced gamblers face two options with very similar expected return and overwhelmingly choose the slightly worse option. They are 1. standing instead of hitting when the player's cards totaling 16 points and the dealer's totaling 10 points (16 vs 10 situations), 2. taking insurance side bets when they have blackjack and 3. taking insurance side bets when they have 'good hands', i.e. card compositions that imply a large probability of winning. For simplicity, we make the approximation that the expected returns of both options are exactly equivalent in all the scenarios.

*3.1.1: blackjack gambling: 16 vs 10 situations*

In this problem, the player may choose to hit (i.e. taking an extra card) or to stand (not taking an extra card). Without any form of restructuring of the problem, the options of hitting and standing are indistinguishable, since for both options, the possible outcomes are equivalent (winning: $\$x$; losing: $\$-x$, where $x > 0$ is the bet size). With the additional constraint that the expected return for both options is the same, it implies that the probability of obtaining those outcomes is also the same. Hence, the preference for standing cannot be predicted by conventional models, including EUT, PT and RT.

So, how can we understand the observed preference for standing through sequential branching? Note that if the player chooses to stand, the player phase is immediately over. The outcome solely depends on what happens in the dealer phase. On the other hand, if the player chooses to hit, the expected returns changes depending on the results of the player phase. If the player busts, the game ends immediately (the player loses). If the player does not bust, the final outcome is still undecided as the dealer has yet to play, but the expected return has now risen because the possibility that the player loses due to busting has been eliminated. The branching schemes for standing and hitting are illustrated in Figure 3.

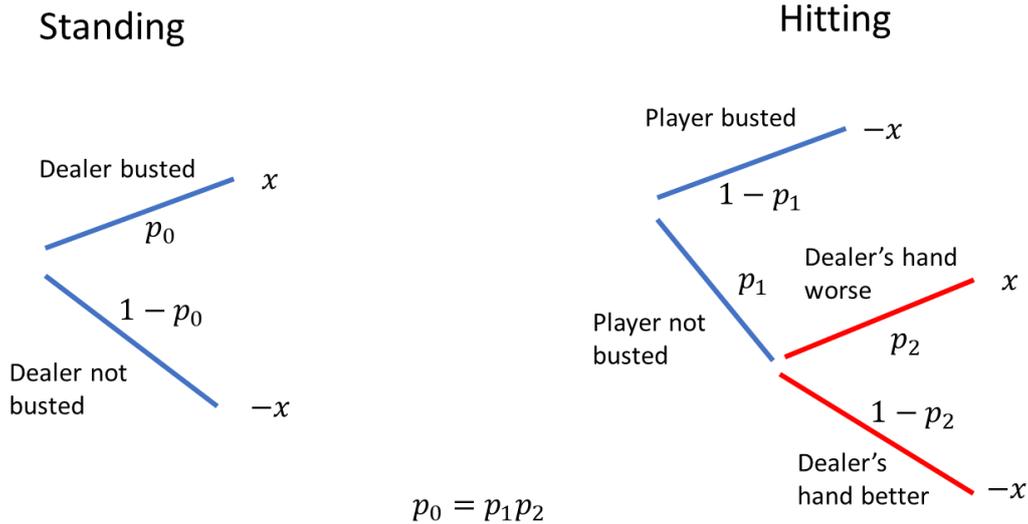

Figure 3: Branching schemes when the player decides to stand (left) and hit (right). Here $p_0$ is the probability that the dealer goes bust, $p_1$ is the probability that the player does not go bust after taking a card, $p_2$ ($0 < p_2 < 1$) is the probability that the player wins the hand (either because their hand is larger, or the dealer goes bust) provided that they do not go bust. The constraint that the expected value for both options is the same means that $p_0 = p_1 p_2$. We further assume that when the player hits, only one more card will be taken, since taking more cards would lead to significantly worse expected return.

The surprise values for standing ($\Delta_{stand}$) and hitting ($\Delta_{hit}$) are computed in Appendix 3. It turns out that $\Delta_{stand} > \Delta_{hit}$ (Please see Appendix 3 for the mathematical details.) This result matches the observed behaviors of gamblers.

This example illustrates how branching can influence the preference of people. In the case of hitting, if the player does not bust, their chance of winning increases, meaning that the intermediate state has a higher expected value than the initial state. Thus, the anticipated surprise for winning is split into two small components: one associated with the transition from the initial state to the intermediate state, and the other with the transition from the intermediate state to the outcome state. Because of the convexity of the surprise function, the aggravation of these small components of surprise is less than the big surprise acquired when the player anticipates standing and winning. By similar reasoning, the negative surprise acquired when the player anticipates hitting and losing is more severe than the case of standing and losing.

As a side note, both PT and RT do not offer any guideline of how to cope with the branching schemes shown in Figure 3 without simplifying it by combining the losing outcomes for hitting, reverting it to the case where both options are indistinguishable.

*3.1.2: blackjack gambling: taking side bets when the player has a good hand*

In this problem, the player may choose to either take or not take an insurance side bet (i.e. betting that the dealer has blackjack. Please refer to Appendix 3 for details of the rules).

*3.1.2.1: Special case: when the player has blackjack*

When the player has blackjack, taking the side bet leads to a certain reward of the bet size, while not taking the side bet leads to a reward of 1.5 times the bet size with the probability $\frac{2}{3}$, and 0 otherwise. Note that this is exactly the same decision-making problem as the one depicted in Table 1 with $p = 1$, $p' = \frac{2}{3}$. We have already established that the certain option, i.e. taking the side bet, is preferred. Sequential branching is irrelevant in this case.

*3.1.2.2: When the player has a general good hand*

The study by Bennis (2004) took place in the US. In most casinos in the US, the dealer peeks for blackjack, which means that before the player phase, the dealer checks if their hand is blackjack. If so, the round promptly ends. If not, the expected return for the player increases, since the original expectation has factored in the possibility that the dealer has blackjack and that has been eliminated. This creates a similar branching scheme as in the 16 vs 10 situations with the player hitting, as depicted in Figure 4.

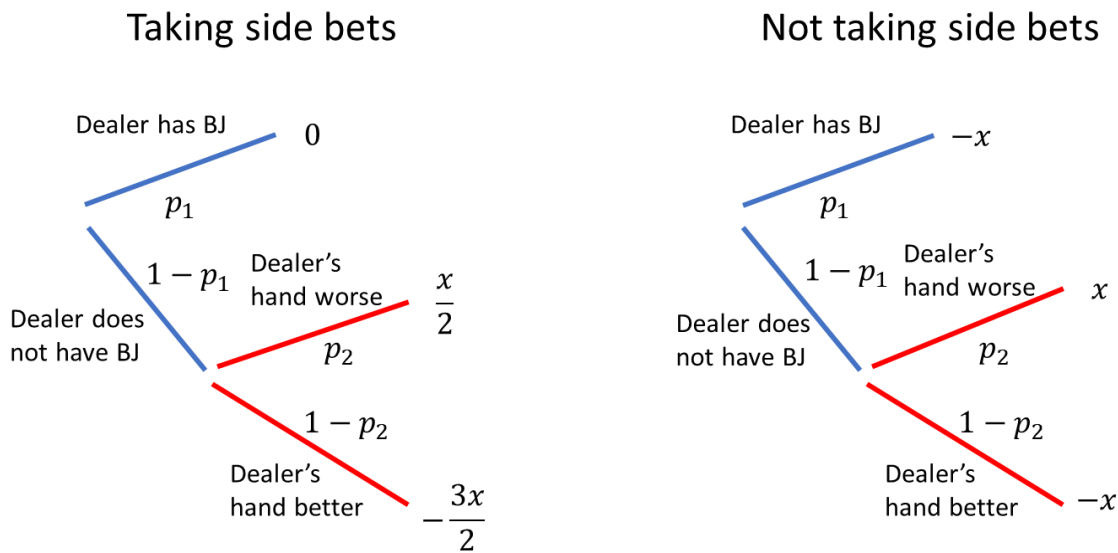

Figure 4: Branching schemes when the player decides to take the side bet (left) and not to take the side bet (right). Here $p_1$ is the probability that the dealer has blackjack, $p_2$ is the probability that the player wins the hand (either because their hand is larger, or the dealer goes bust) provided that the dealer does not have blackjack.

Since the risk aversion factor $k$ only appears when negative surprise is incurred, the expressions for $\Delta_{bet}$ and $\Delta_{no\ bet}$ depend on the value of $p_2$. We will leave the mathematical details to Appendix 3 and numerical simulation results to Figure S3 in Appendix 5. Here, we state the results that $\Delta_{bet} > \Delta_{no\ bet}$ if $p_2 > \frac{3}{8}$ (Note that the converse is not true. The result is inconclusive if $p_2 \leq \frac{3}{8}$). Since a large $p_2$ corresponds to a 'good hand', the prediction of the model matches the observed behaviors that people choose to take the side bet when they have a good hand.

To understand the results, first note that the red branches for both options are effectively equivalent, since the deviations from the updated expected values for the outcomes are the same. It makes sense since the side bet is already lost when it is revealed that the dealer does not have blackjack, while the red branch corresponds to the situation after the above-mentioned revelation. For the blue branches, taking the side bet leads to relatively 'certain' outcomes. With good hands, the player's chance of winning is large enough such that the decision-making problem is outside the regime of lottery. In such cases, as we have studied in Section 1, risk aversion dominates, and certain outcomes are preferred.

In Appendix 5, we show that the observed behavior cannot be explained by both PT and RT.

*3.1.4: final remarks*

One may argue that these decision-making problems are so complex that the gamblers misinterpret the probability of winning underlying each option. However, information on the 'optimal strategy', i.e. the best option to take in order to maximize the expected return, for each initial card composition is well known and widely available in public. Bennis (2004) showed that experienced gamblers are indeed familiar with such optimal strategy, suggesting that gamblers are mostly likely aware that the options they choose are slightly inferior (again, in terms of expected return).

## 3.2 Ambiguity: Ellsberg paradox (2-urn problem)

In the above examples, the intermediate states (e.g., if the dealer has blackjack or not) are revealed during the game. Here, we propose that this identifiability of intermediate states is not necessary for sequential branching to happen. The intermediate states could remain unidentified and just be a mental product of people conceptualizing a problem. To illustrate this, let us consider the 2-urn version of the Ellsberg paradox (Ellsberg, 1961). In the experimental set-up, the subjects are told that there are two urns: one containing $n$ red balls and $n$ black balls; the other containing an undisclosed number of red and black balls totaling $2n$. the subjects will then pick a ball in one of the urns. They will win $1 if the ball is red (or black, the color does not matter in the observed behavior) and win nothing otherwise. It has been shown that most subjects prefer to pick balls from the urn with known ball composition. This phenomenon is known as ambiguity aversion.

Let us call picking balls from the urn with known composition 'Option 1' and picking balls from the other urn 'Option 2'. For Option 1, it gives 50% chance of a reward of $1 and 50% of a reward of $0. The surprise value $\Delta_1$ is given by:

$$\Delta_1 = \frac{1}{2}\delta\left(\frac{1}{2}\right) + \frac{1}{2}\delta\left(-\frac{1}{2}\right) \tag{6a}$$

For Option 2, it has been proposed that subjects may view it as a compound lottery, i.e. they first envision the various possible compositions in the urn (Krähmer & Stone, 2013; Segal, 1987). In the context of our model, these possible urn compositions constitute the intermediate states. The intermediate states can be parameterized by the fraction $m$ ($m = 0, \frac{1}{2n}, \frac{2}{2n}, \dots, 1$) of balls that are red. Without further information, one may assume that the probability distribution underlying the ball compositions is symmetric about even number of red and black balls, i.e. the probability for entering the intermediate states, $p$, is constraint by $p(m) = p(1 - m)$ for all $0 \leq m < \frac{1}{2}$. For each intermediate state $m$, there are two final outcomes ($1 with probability $m$ or 0 with probability $1 - m$). Under this formalism, the surprise value $\Delta_2$ for Option 2 is given by

$$\Delta_2 = \sum_m p(m) \left[ \delta\left(m - \frac{1}{2}\right) + m\delta(1 - m) + (1 - m)\delta(-m) \right] \tag{6b}$$

Then, the difference is computed as

$$\Delta_1 - \Delta_2 = (k - 1) \sum_{m<1/2} p(m) \left[ f\left(\frac{1}{2} - m\right) + mf(1 - m) + (1 - m)f(m) - f\left(\frac{1}{2}\right) \right] \tag{7}$$

In eq. (7), any symmetric pair of intermediate states $m$ and $1 - m$ gives the same magnitude of surprise with opposite signs, and therefore, the positive perfector $k - 1$ appears after combining their influence (See Appendix 4 for the mathematical details). $\Delta_1$ can be larger or smaller than $\Delta_2$ depending on the choice of $f$, suggesting that both ambiguity seeking and ambiguity aversion is in principle possible. However, we can show that ambiguity aversion is guaranteed if $f$ is not very convex (e.g., $f''(m) \leq \frac{3}{2} f'(m)$ for $0 < m < 1/2$) or if $f$ is highly convex (e.g., $\mu_0 f(1 - \mu_1) > f\left(\frac{1}{2}\right)$, with $\mu_c = \frac{\sum_{m<1/2} m^{c+1} p(m)}{\sum_{m<1/2} m^c p(m)}$, which is between 0 and 1/2; see Appendix 4). Figure 5 plots $\Delta_2 - \Delta_1$ using power functions, $f(m) = m^r$, assuming a uniformly distributed $p(m)$. Consistent with our analysis, ambiguity aversion is observed both at small and large $r$, corresponding to the regime where $f$ is mildly and strongly convex. In this case, ambiguity aversion dominates over a large parameter space, which echoes the general preference for Option 1 as observed in experiments.

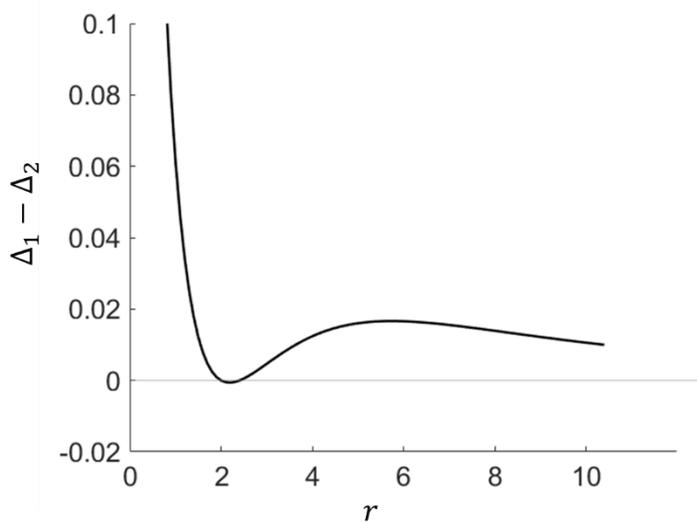

Figure 5: $\Delta_1 - \Delta_2$ vs $r$ for $f(m) = m^r$ and $p(m) = \frac{1}{2n+1}$. $\Delta_1 - \Delta_2 > 0$ (ambiguity aversion) except for the small regime $2 < r < 2.5$. $n = 50, k = 2$.

To understand the results, imagine that during the thought process in evaluating the value of picking the ambiguous urn, we anticipate that the ambiguous urn may contain more balls with the prize-winning color than our initial expectation. We would be pleasantly surprised by this potential scenario, culminating in a positive surprise in the intermediate state. Nevertheless, in the 2$^{nd}$ branch, with the expectation now risen, we would have a negative anticipated surprise going from the intermediate state to the final state because of the possibility of not winning, which would lead to an outcome much worse than the updated expectation (The 2$^{nd}$ branch is exactly the problems we discussed in Section 1. In this scenario, $p$ is large). In the end, it is a trade-off between the surprise generated from the first and that from the second transition. The outcome of this trade-off determines one's affinity for ambiguity.

As we mentioned, we do not have to assume that the subjects are actually informed of the urn composition and/or the underlying probability distribution of the composition for the effects of sequential branching to kick in. How any potential difference in behavior when they are told of the urn composition and/or their distribution, and when they are not, can be theorized, while interesting, is not within the scope of this work (For experimental work, please refer to e.g. Halevy (2007)).

One may argue that when $n$ is large, it is infeasible to mentally consider the huge number of branches as depicted in Figure 2. In practice, people may evaluate only a limited number of intermediate states such as a single pair of symmetric states, $m$ and $1 - m$. As shown above as well as in Appendix 4, ambiguity aversion holds robustly in this case (regardless of the form of $p(m)$) if $f$ is not very convex (or if $f$ is highly convex). Similarly, people may use coarse-grained intermediate states to approximate the surprise value.

As a final note, like the blackjack 16 vs 10 problem in Section 3.1.1, the difference in the branching structure between the options means that PT and RT cannot account for the behavior.

## 3.3 Segregation of probable and improbable outcomes

Now we turn to another type of scenarios where sequential branching could possibly occur. Here, intermediate states do not correspond to physical states but reflect mental representations that group multiple improbable outcomes together instead. In real life, there are many events that we may encounter but with very low probability, like earthquake, traffic accident, winning a jackpot. While we do not completely ignore these events, they are not processed in conjunction with other more probable events. For example, we may buy insurance or draw separate contingency plans to mitigate the effects for several disastrous events at once. After that, we may not consider these events during our daily activities. In the context of our model, branches comprising of all these rare events are created. To illustrate this, let us consider the original version of Allais paradox (Allais, 1953). The decision-making problems in Allais paradox are shown in Table 2.

(a) Problem 1

|  | Option 1 | | Option 2 * | |
|---|---|---|---|---|
|  | Reward | Probability | Reward | Probability |
| Outcome 1 | 0 | 0.89 | 0 | 0.9 |
| Outcome 2 | 1 | 0.11 | 5 | 0.1 |

(b) Problem 2

|  | Option 1 * | | Option 2 | |
|---|---|---|---|---|
|  | Reward | Probability | Reward | Probability |
| Outcome 1 | 1 | 1 | 0 | 0.01 |
| Outcome 2 |  |  | 1 | 0.89 |
| Outcome 3 |  |  | 5 | 0.1 |

Table 2: The two decision-making problems in the Allais paradox. The asterisk depicts the option preferred by most people.

In Problem 1, Option 2 is preferred by most people. This can easily be understood by the significantly larger reward size for Option 2 as compared to the marginally smaller probability in obtaining the reward. Problem 2 is obtained by replacing the probability 0.89 portion of the reward '0' by reward '1' for both options. Despite this equal treatment of both options, Option 1 is favored in Problem 2 instead. The two options have slightly different expected return, but we assume that it plays a relatively minor role in decision-making and therefore ignore its effect (but see Discussion for a potential extension to model problems with options having different expected values). First, we will show how sequential branching, by splitting the improbable outcomes, '0' and '5' in Problem 2, into sub-branches, makes Option 2 less appealing than the case without the splitting. The branching schemes for Option 2 when the improbable outcomes are grouped together, and when they are not, are depicted in Figure 6.

(a)

No grouping for improbable outcomes

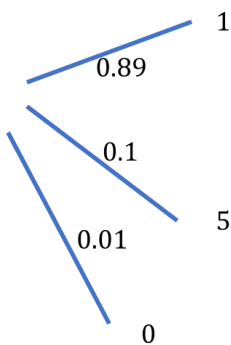

With grouping for improbable outcomes

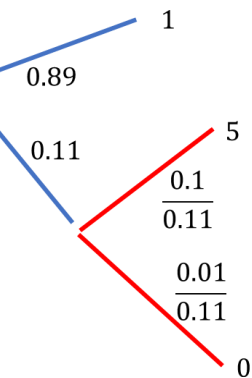

(b)

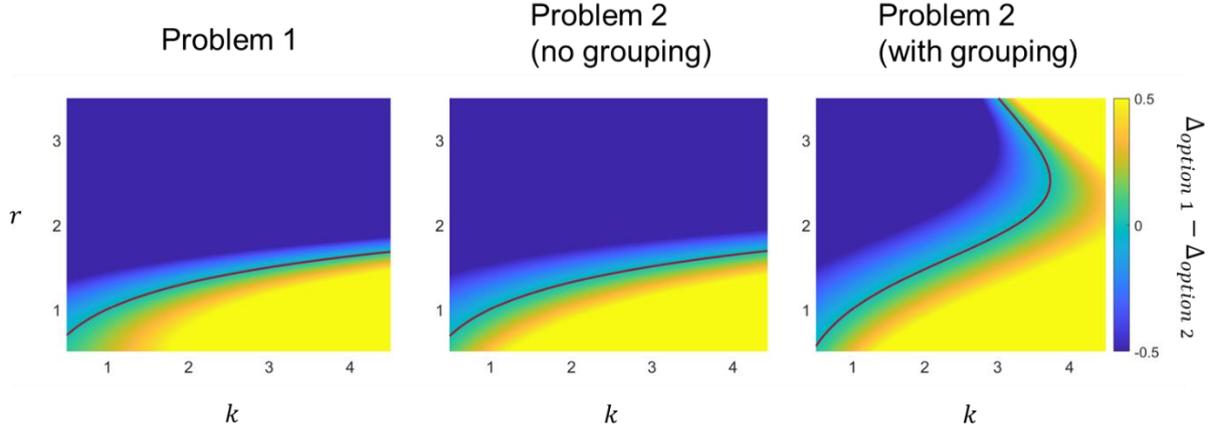

Figure 6: (a) The branching schemes for Option 2 of Problem 2 in Table 2 when the improbable outcomes are grouped together and when they are not. (b) The difference in surprise values between Option 1 and Option 2 for problem 1 (left), Problem 2 without grouping the improbable outcomes (middle), Problem 2 with the improbable outcomes grouped (right). Yellow (Blue) color corresponds to the regime where Option 1 (2) is preferred. The dark red line is the boundary where the options are equally preferred, i.e. $\Delta_{option\ 1} = \Delta_{option\ 2}$. We set $f(x) = x^r$.

The surprise value with (and without) grouping $\Delta_{group}$ (and $\Delta_{no\ group}$) are given by:

$$\Delta_{group} = -0.89kf(E_0 - 1) + 0.1\big(f(E_1 - E_0) + f(5 - E_1)\big) + 0.01\big(f(E_1 - E_0) - kf(E_1)\big) \tag{8a}$$

$$\Delta_{no\ group} = -0.89kf(E_0 - 1) + 0.1f(5 - E_0) - 0.01kf(E_0) \tag{8b}$$

with the expected values at the initial state $E_0 = 0.89 * 1 + 0.1 * 5 = 1.39$ and intermediate state $E_1 = (0.1/0.11) * 5 = 4.55$. In eqs. (8a) and (8b), the first, second, and third terms in the right-hand side represent the contributions from obtaining the outcomes '1', '5', and '0', respectively. As the result of grouping, the positive surprise of getting the outcome '5' is reduced (since $f(E_1 - E_0) + f(5 - E_1) < f(5 - E_0)$ for any convex and increasing $f$) and the negative surprise of getting the outcome '0' is magnified (since $kf(E_1) > kf(E_0) + f(E_1 - E_0)$ for any convex and increasing $f$ and $k \geq 1$). This means that with grouping, Option 2 is more unfavorable than the case without, implying that the parameter space where Option 2 is worse than Option 1 becomes larger when branching is considered.

Now, by using a common class of convex functions $f(x) = x^r$, we compare the surprise values of the two options for both problems numerically. As shown in Figure 6b, for Problem 1, Option 2 is preferred for most regimes (only except when $f$ is weakly convex and the negative surprise is exacerbated by large $k$) as predicted. For Problem 2, without considering grouping, the preference remains unchanged in contrary to the experimental observation. However, when grouping is considered, in line with our analysis, the surprise value for Option 2 reduces (See also Figure S4) such that Option 1 is now preferred for a sizable parameter space. Our model is consistent with the experimental observation either when $k$ is not too large ($k < 3$) and $f$ is moderately convex ($1 < r < 2$), or when $k$ is very large and $f$ is very convex.

Please note that Allais paradox can also be reproduced by CPT and RT (See Appendix 4). In fact, Problem 3, 4, 7, 8 in Kahneman & Tversky (1979) is sometimes considered as a variant of Allais paradox. However, our model explains them using different mechanisms: Problem 3, 4, 7, 8 in Kahneman & Tversky (1979) by the convexity of the surprise function and the original version of Allais Paradox by grouping of improbable outcomes.

Outcome grouping is relevant for a wide range of decision-making problems beyond variants of Allais paradox, for instance when there are multiple events that give the same outcome. In EUT and CPT, such events are effectively combined into a single event. In other words, these models predict that experimentally, combining same-outcome events makes no difference in the observed behavior. However, this is not necessarily true, as illustrated by an experiment in Birnbaum (2008, Table 1) shown in Table 3.

(a) Problem 1

| Option 1 | | | Option 2 * | | |
|---|---|---|---|---|---|
| Color of ball drawn | Reward | Probability | Color of ball drawn | Reward | Probability |
| Red | 100 | 0.85 | Black | 100 | 0.85 |
| White | 50 | 0.1 | Yellow | 100 | 0.1 |
| Blue | 50 | 0.05 | Purple | 7 | 0.05 |

(b) Problem 2

| Option 1 * | | | Option 2 | | |
|---|---|---|---|---|---|
| Color of ball drawn | Reward | Probability | Color of ball drawn | Reward | Probability |
| Black | 100 | 0.85 | Red | 100 | 0.95 |
| Yellow | 50 | 0.15 | White | 7 | 0.05 |

Table 3: The two decision-making problems in Birnbaum (2008, Table 1). The asterisk depicts the option preferred by most people.

Although the two problems are identical if the same-outcome events are combined, most people choose different options for Problem 1 and Problem 2. This motivates us to again investigate the branching schemes that group the improbable events and without combining the same-outcome events, as shown in Figure 7a. Again, we neglect the effects of different expected returns for the two options (c.f. Discussion).

(a)

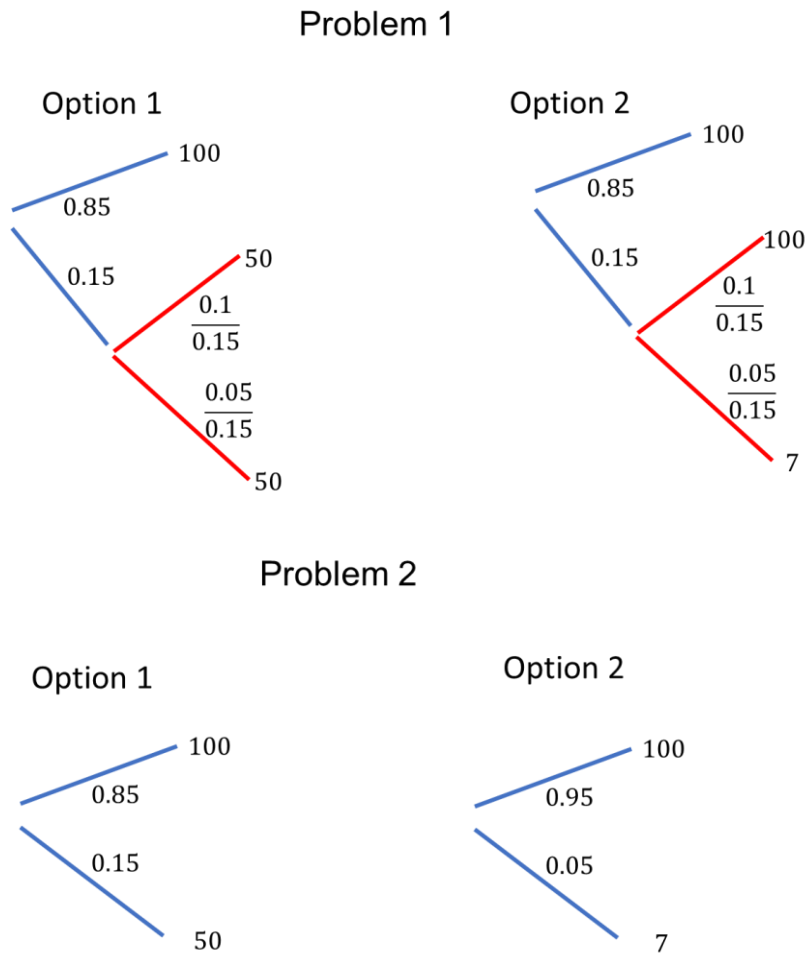

(b)

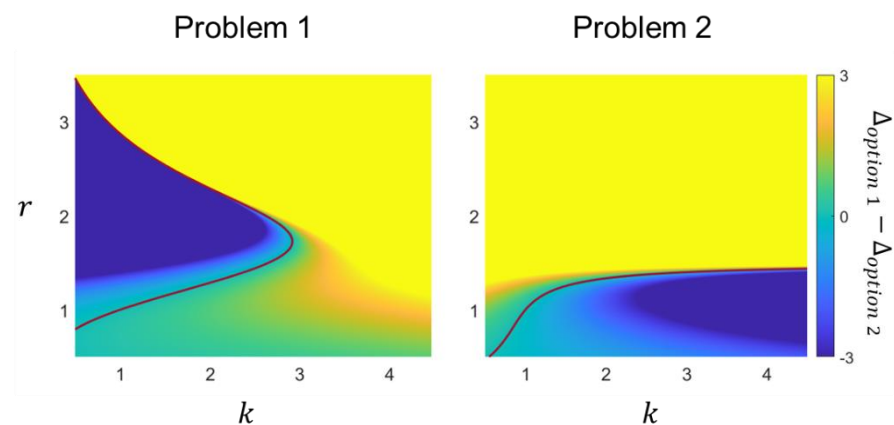

Figure 7: (a) The branching scheme for Problem 1 (top) and Problem 2 (bottom) in Table 3. (b) The difference in surprise values between Option 1 and Option 2 for Problem 1 (left), Problem 2 (right). As in Figure 6b, yellow (Blue) color corresponds to the regime where Option 1 (2) is preferred. The dark red line is the boundary where the options are equally preferred, i.e. $\Delta_{option\ 1} = \Delta_{option\ 2}$. we set $f(x) = x^r$.

For Option 1 in Problem 1, the events that lead to the same outcome are both improbable, and therefore get grouped into the same branch. In this case, they can be combined since the branch they are in leads to a certain outcome, i.e. no 'surprise'. In other words, Option 1 of Problem 1 and 2 are equivalent. For Option 2 in Problem 1, one event with the reward of '100' is probable while the other is improbable, causing them to be grouped into separate branches. In this case, the branches cannot be combined (Note the similarity to Figure 6a).

The surprise values for Option 2 in Problem 1 ($\Delta_{group}$) and Problem 2 ($\Delta_{no\ group}$) are given by

$$\Delta_{group} = 0.85f(100 - E_0) - 0.15kf(E_0 - E_1) + 0.1f(100 - E_1) - 0.05kf(E_1 - 7) \quad (9a)$$
$$\Delta_{no\ group} = 0.95f(100 - E_0) - 0.05kf(E_0 - 7) \quad (9b)$$

with the expected values at the initial state $E_0 = 0.95 * 100 + 0.05 * 7 = 95.35$ and intermediate state $E_1 = \frac{0.1}{0.15}100 + \frac{0.05}{0.15}7 = 69$. For our model to explain the experimental results, $\Delta_{group} > \Delta_{no\ group}$ should hold. This is true if $k = 1$ due to the convexity of the surprise function $f$ in the gain domain, similar to the case in the Allais paradox. More generally, when $k > 1$, $\Delta_{group} > \Delta_{no\ group}$ requires that the effects of $f$'s convexity overwhelm that of the risk aversion factor $k$. We illustrate this by comparing the surprise values of the two options for both problems numerically. As the consequence of event grouping, the surprise value for Option 2 increases as long as $r$ is not too small (See Figure S4), such that the model can reproduce the experimentally observed preference when $k$ is not too large ($k < 3$) and $f$ is moderately convex ($1 < r < 3$) (Figure 7b). Note that this regime roughly overlaps with the regime where our model's prediction is consistent with experimental observation for the Allais paradox we previously discussed (at roughly $2 < k < 3$ and $1.2 < r < 1.8$).

In both the Allais paradox and the Birnbaum problems, the preferred options can be altered by grouping improbable events together and postponing the detailed anticipation of each individual event to a later branching step. As a result, the associated surprise is broken down into a general one about the group and a specific one about individual events in the group. Since the surprise function is nonlinear, it results in changes in the total aggregated surprise.

## Section 4: Discussion

In this work, we introduce a decision-making model based on anticipated surprise that tackles problems in which the predictions from EUT are inconsistent with experimental observations. The model hinges on 3 main assumptions: 1. The reference point being the expected value of all outcomes; 2. The surprise function is convex in the gain domain (and concave in the loss domain) and 3. general bias towards risk aversion ($k > 1$). Our model can explain Problem 3, 4, 7, 8 in Kahneman & Tversky (1979) without having to introduce extra concepts, like subjective probability weighting as PT does.

We then introduce the idea of updating the reference point along the sequential stages of a problem during the anticipation process. We outline 3 scenarios in which sequential anticipation is applicable: 1. revelation of additional information about the outcome; 2. ambiguity and compound lottery; 3. Segregation of probable and improbable outcomes. The sequential anticipation model can predict behaviors that cannot be explained by EUT, PT and RT, especially those involving options that are considered equivalent by those models. Our numerical results and mathematical analysis show that in order to replicate the experiments, the surprise function needs to be moderately convex, and $k$ needs to be larger than 1 but not too large. This is generally in line with the assumption of the model we made when we considered the single step problems in Section 1.

Our sequential anticipation formalism allows us to have new perspectives on the axioms of EUT and the editing rules of PT, e.g. change of reference points, combination of events that lead to the same outcomes providing guidelines of when they are valid and when they are not, and why, as demonstrated in our discussions on Problem 10-12 in Kahneman's 1979 paper and the examples in Section 3. Furthermore, as shown in the examples we presented, decisions in multi-stage problems are often mostly based on a particular dominant branch as the result the convexity of the surprise function. This constitutes an experimentally testable prediction, since given a problem with a particular branching structure, the model can identify the branches that are responsible for the observed behavior. By doing so, one may be able to reduce the problem structure to a simpler one. For example, the model suggests that the Blackjack side bets problem (See Section 3.1.2.2, Figure 4) can be simplified to a corresponding single-stage problem by removing the red branches and replacing them with their expected values.

## 4.1 Difference between our model and models related to the prospect theory

The most striking difference between our model and PT involves the mechanism for reproducing people's opposite risk preference for gambles with small and large probability of winning/losing ($p$) observed in Problem 3, 4, 7, 8 in Kahneman & Tversky (1979). The fundamental reason why EUT cannot reproduce the experimental observations in these problems is that with the reference point being held at the baseline level, convexity (or concavity) of the utility function $u$ alone can only lead to risk-seeking or risk-averse behavior irrespective of $p$. PT and CPT tackle this problem by introducing nonlinear probability weighting. More specifically, these models use a concave (convex) $v$ in eq. (1) for the gain (loss) domain. This property alone leads to risk-averse (risk-seeking) behaviors, which matches the behaviors observed for large $p$. In order to explain the opposite behavior when $p$ is small, nonlinear probability weighting is used to overwhelm the effects of the concavity/convexity of $v$. A similar logic applies to models that implicitly introduce effective probability weighting. For example, the model by Gul (1991) could reproduce the risk-seeking behavior at small $p$ in the gain domain by using a convex $v$, but it has to resort to 'disappointment aversion', which mathematically corresponds to probability weighting, to reproduce the risk-averse behavior at large $p$. (The model needs further assumptions to account for the risk preference in the loss domain.) By contrast, through using the expected value of the outcomes as the reference point, our proposed model can reproduce the risk preference at both large and small $p$ by mere convexity (concavity) of $\delta$ in the gain (loss) domain. Another issue with nonlinear probability weighting is that, as discussed in Introduction, subjective probability perception is highly context-dependent (Gallistel et al., 2014; Hertwig et al., 2004; Wu et al., 2009; Zhang & Maloney, 2012), meaning that there is a risk that the nonlinear probability weighting is overfitted to the specific problems studied.

In order to account for some experimental observations associated with the framing effect, PT and CPT introduce several editing rules, which change the reference point and modify some outcomes by splitting and/or disregarding them (Kahneman & Tversky, 1979). While they have provided guidelines on how to apply these rules depending on the structure of the problem, a large degree of freedom is left to the user of the model for the disposition of the rules. There are situations, e.g. in the original version of Allais paradox (Allais, 1953), where multiple editing rules are eligible according to those guidelines. The choice of which rules to apply and/or the order of their application could lead to different predictions of behaviors (Birnbaum, 2008). In addition, as we mentioned in Section 2.1, in the face of problems with sequential stages, PT still uses a single, static reference point, which allows the conversion of the problem into a single-stage one. This simplification inherently assumes that people's expectation does not notably change when they process through different sequential stages mentally. If the reference point is deemed to have altered in some sequential stages, the problem structure cannot be collapsed into single-stage using existing mechanisms in PT, and because of this, it is unclear how such problems should be handled by PT. On the other hand, our proposed model has a clearly defined reference point: the expected value of the outcomes at the next branching step. The mechanism of sequential branching not only provides an alternative perspective on issues surrounding these editing rules and resolves some of the ambiguity involved (See e.g. Section 3.3), but also naturally models the dynamically changing reference points during decision-making processes. One may ask if similar sequential branching mechanism could be introduced to PT and CPT. However, as PT and CPT use nonlinear probability weighting, it is non-trivial to formulate Markovian transitions of intermediate states under their framework.

A possible weakness of our model as compared to PT is that it predicts excessive risk-averse behavior in the loss domain. For example, it leads to the prediction that people would take the more certain options for both the gain and loss regime at around $p = 0.5$ (Figure 1) when $k > 1$, which could be at odds with some experimental results (Kahneman & Tversky, 1979). Nevertheless, generally speaking, preferences of people tend to be ambiguous at intermediate values of $p$ (Ruggeri et al., 2020; Somasundaram & Diecidue, 2017). So, qualitatively speaking, the partial lack of reflection effect predicted by our model in this regime may not be a fundamental weakness of the model. Another related problem is the 'Asian disease problem' (Tversky & Kahneman, 1981). The experiment showed that people essentially change their choices from being risk-averse to being risk-seeking when the description of options emphasizes the loss instead of the gain in the possible outcomes. A common way of formulating the problem is by shifting all outcomes by a constant value as shown in Tversky & Kahneman (1981), which produces the same preference as the classical 'Asian disease problem'. Our model in its current form cannot reproduce the correct risk choice because anticipated surprise is computed only in relative to the expected outcome, which also shifts along with the individual outcomes. However, there are multiple possible mechanisms to modify our model such that it could explain the risk-seeking behaviors in the loss domain. For example, one could make $k$ (the risk aversion factor) a variable which is decreasing with the expected outcome, so that as the expected outcome becomes negative, $k$ becomes small, and hence the decision maker becomes more risk-seeking. Alternatively, the reference point might be slightly reduced from the expected outcome in the loss domain to alleviate the disappointment when getting a bad outcome and boost the surprise when getting a good outcome in a gamble. In a word, these problems alone do not refute the basic hypothesis of our model that expected outcome determines the reference point for economic decision-making problems.

## 4.2 Difference between our model and the regret theory

With the convex surprise function and the lack of probability weighting, one may find resemblance between our model and the regret theory (RT) (Bleichrodt & Wakker, 2015; Loomes & Sugden, 1982). In RT, decision-making is based on 'anticipated regret', computed by the difference in value between the outcome of an action and the outcome should other actions be chosen. RT can also reproduce Problem 3, 4, 7, 8 in Kahneman & Tversky (1979) (See e.g. eq. (7) in Loomes & Sugden (1982)). Nevertheless, there are fundamental differences between RT and our model. Unlike RT, the evaluation of an option is irrelevant to the evaluation of other options in our model. Here, we would like to emphasize that 'regret' and 'surprise' are not competitive in nature, and they may provide us with complementary views on how a decision is made. It is plausible that for some problems, both 'anticipated regret' and 'anticipated surprise' play a part in decision-making. However, we also note that it may be difficult to evaluate anticipated regret when the outcome structures of the options available are very different, since in such situations, there is no apparent way to conceive how the outcomes among these options would correspond to each other. This problem is aggravated when we consider sequential anticipation, where the number of plausible ways to relate intermediate states of different options explode as the number of branches increases. In such cases, regret may be too fuzzy to be perceived even after a decision has been made and outcome received, let alone anticipating it beforehand.

## 4.3 Decision-making among options with different expected values

In this work, we are focusing on decisions among options with similar expected outcomes. However, in real life, most decision-making problems involve options with significantly different expected outcomes. It is therefore of interest to extend our model to also cover these cases. A possible way is to consider the change from the status quo to the initial expected value of an option to be part of the sequential anticipation process, i.e. to affix initial expected value $E(t = 0) = 0$ as the status quo in eq. (4) such that the next step corresponds to the initial state of the decision problem with $E(t = 1)$ being the expected value of all outcomes. This extension naturally interpolates between the basic decision-making scenario solely based on expected outcomes that EUT targets and the scenario solely based on risk-based surprise that our current work targets. One caveat of this approach is that in principle, the influence of surprise could still unrealistically dominate over the influence of the expected values in the mixed domain (in which possible outcomes include both gain and loss) if the effect of an outcome-outlier is overly magnified by a highly convex $f$ and/or large $k$. However, this is unlikely if $f$ is only mildly convex and $k$ is not too large, which is consistent with the parameter regime suggested by the examples examined in this work. This proposed transition from the baseline level to the initial state based on the expected value of an option is also biologically plausible. Dopamine neurons are essential for action execution and encoding reward prediction errors (DeWitt, 2014; Schultz, 2010). In conditioning experiments, these neurons give a first response when a reward prediction signal is given, which reflects the difference between the expected reward associated with the signal and the baseline level, and then give a second response when the reward is given, which reflects the difference between the given and the expected reward (Pan, Schmidt, Wickens, & Hyland, 2005; Schultz, 2010). In the context of our model, a similar sequence could be anticipated during planning with the first component of the anticipated surprise being driven by a (mentally) simulation of picking an option and the second component being driven by a simulation of outcome presentation.

## 4.4 Possible neural implementation of anticipated surprise

Decision-making is often studied in neuroscience. Our decision-making model is based on two concepts: 'surprise' and 'anticipation'. Both concepts have strong neural bases. For the former, as discussed in the previous section, the activities of dopamine neurons encode the reward prediction errors, i.e. the deviation of the actual reward from the expectation of the experimental subject (DeWitt, 2014; Schultz, 2010, 2017). Stauffer, Lak, & Schultz (2014) showed that the above-mentioned expectation can be well approximated by the expected value of the reward.

For the latter, it has been shown that shortly before an animal executes an action, hippocampus place cells exhibit a sequence of firing patterns that are highly predictive of the sequence of action the subject takes later (Ólafsdóttir et al., 2018; Pfeiffer & Foster, 2013). This suggests that these anticipative patterns may be relevant to planning and deciding what actions to take by the subject. These neural activities related to anticipation is observed in multiple brain areas, including the hippocampus, the neocortex, and the dopaminergic midbrain. Integrated information from these areas is predictive of the decisions made by people (Iigaya et al., 2020).

Another question is how the evaluation in different sequential branches depicted in our model could be efficiently computed in neural systems, especially for complex problems with an elaborated branching structure. A recent work (Dabney et al., 2020) presented findings that are suggestive of parallel computing by dopamine neurons. In their experiment, subjects are given a cue representing probabilistic reward, and it was shown that a subset of dopamine neurons exhibit different reversal points characterizing the reward-activity curve. Also, some neurons only respond strongly to positive surprise and some only to negative surprise. The variety in their response profile allows them to simultaneously encode the surprise associated with probabilistic outcomes. Parallel computing during decision-making could have practical importance as suggested by experimental and theoretical studies. Parallel computing implemented in brain areas relevant for planning and action selection, like the basal ganglia (Alexander & Crutcher, 1990), enhances the efficiency of searches through branching trees (Kriener, Chaudhuri, & Fiete, 2020), which is especially important in situations where decisions have to be made under time pressure (Wan et al., 2011).

Our theory of how risk-preference is computed from anticipated surprise may provide a novel direction for exploring the neuroscience bases of decision-making in economics that involves branching outcomes.

## 4.5 Effect of learning on decision-making under anticipated surprise

The problems we studied are hypothetical in the sense that the probability and the size of rewards are explicitly given (Blackjack could be considered an exception, though the probability of winning for each card combination and for each option to take is widely available). In real life, these elements may have to be learnt through trial-and-error. There is ample evidence that learning and past experiences affect decision-making. Momennejad, Otto, Daw, & Norman (2018) shows that offline replay, corresponding to learning and memory consolidation, alters the subsequent decision made during tasks. Chen, Kim, Nofsinger, & Rui (2007), and Hoffmann & Post (2017) shows that a trader's return in the recent past affects their risk attitude and expectation on future return, even though it is often unclear whether past return is indicative of future return due to the volatility of the financial market.

How this learning is implemented in conjunction with decision-making under risk is outside of the scope of the model in this work. Simplistically, one could use one of Bayesian reinforcement learning frameworks to train an "anticipated surprise model" for risky decision-making. Nevertheless, it is unlikely that learning and decision-making can be completely decoupled. Based on the similarity of online replay (during task performance) and offline replay (outside task performance), it is strongly suggested that that similar neural activity patterns are involved in planning and learning (Ólafsdóttir et al., 2018). Moreover, anticipated surprise likely changes along the learning stages. For example, it has been shown (Burger, Hendriks, Pleeging, & van Ours, 2020) that both the happiness for winning a small price and sadness for losing in a lottery is smaller for people who have bought a ticket than those who have not bought a ticket but rather given one. Since inevitably, there were people who regularly buy tickets, and therefore frequently experienced winning and losing, in the group of ticket buyers, the results could possibly be interpreted as desensitization of emotions by repeated exposure to certain events. Indeed, it has been shown that responses from dopamine neurons, which modulate emotional processing, habituate with repeated presentations of the same stimuli (Ardiel et al., 2016; Menegas, Babayan, Uchida, & Watabe-Uchida, 2017). Given the role of dopamine neurons in relation to decision-making and encoding reward prediction errors, it is not unreasonable to speculate that such repeated stimulation may have a similar desensitizing effect on anticipated surprise. The interplay between learning and decision-making is an interesting future topic to study.

# Acknowledgement

The study was supported by RIKEN Center for Brain Science, Brain/MINDS from AMED under Grant no. JP21dm0207001 and KAKENHI Grant-in-aid JP18H05432.

# Appendix 1: Reproducing the patterns of prospect theory in a model with surprise functions of general form

Here we are showing that the predictions from our model conform to the experimental observation in Kahneman & Tversky (1979) for the type of problems described in Table 1. For the gain domain, it is useful to study how the surprise function varies with respect to $p$ in the setting described in Table 2.

|  | Reward | Probability |
|---|---|---|
| Outcome 1 | $1/p$ | $p$ |
| Outcome 2 | 0 | $1 - p$ |

Table A1: The setting for studying the property of the surprise function in the gain domain. Note that we set $\bar{x} = 1$ to simplify analysis.

In this setting, the surprise value $\Delta$, as a function $p$, is given by:

$$\Delta(p) = p\delta\left(\frac{1}{p} - 1\right) + (1-p)\delta(-1) = pf\left(\frac{1}{p} - 1\right) - (1-p)kf(1) \tag{A1}$$

From (A1), we can obtain the following important results:

$$\Delta(p = 1) = 0$$

$$\lim_{p \to 0} \delta(p) = \lim_{p \to 0} pf\left(\frac{1}{p}\right) = \infty$$

$$\frac{\partial \delta}{\partial p}\bigg|_{p=1} = kf(1) - f'(0) \geq f(1) - f'(0) > 0$$

$$\frac{\partial^2 \delta}{\partial p^2} = \frac{1}{p^3} f''\left(\frac{1}{p} - 1\right) > 0$$

The 2nd and 4th conditions mean that $\Delta(p)$ is positive and decreasing at small $p$. The 1st, 3rd and 4th conditions necessitate that $\Delta(p)$ reaches negative value at some intermediate values of $p$, and then increases and reaches the value of 0 at $p = 1$. This results in a U-shaped $\Delta(p)$ that crosses zero at intermediate $p$. See, for example, figure 1.

The value of $p$ where $\Delta(p) = 0$, aside from the trivial solution $p = 1$, corresponds to the probability where the subject switch from the gambling option to the certain option. For the special case where $k = 1$, it can be easily shown from (A1) that $\Delta(p = 0.5) = 0$. For the realistic case where $k > 1$, note that $\delta$ is decreasing with $k$, which means that $\Delta(p = 0.5) < 0$ and that $\Delta$ crosses 0 at a smaller value of $p$.

In the loss domain, the relevant setting is shown in Table A2.

|  | Reward | Probability |
|---|---|---|
| Outcome 1 | $-1/p$ | $p$ |
| Outcome 2 | 0 | $1 - p$ |

Table A2: The setting for studying the property of the surprise function in the loss domain.

In this setting, the surprise value $\Delta_-$ is given by:

$$\Delta_-(p) = pkf\left(\frac{-1}{p} - (-1)\right) + (1-p)f(-(-1)) = -pkf\left(\frac{1}{p} - 1\right) + (1-p)f(1) \qquad (A2)$$

We are particularly interested in knowing whether reflection effect holds in our model. For a pair options with probability $p_1$ and $p_2$ respectively, reflection effect holds when $\Delta(p_1) > \Delta(p_2)$ and $\Delta_-(p_1) < \Delta_-(p_2)$, or when $\Delta(p_1) < \Delta(p_2)$ and $\Delta_-(p_1) > \Delta_-(p_2)$. For $k = 1$, $\Delta_-(p) = -\Delta(p)$, which implies that the reflection effect can be observed at all values of $p$, since $\Delta(p_1) > \Delta(p_2)$ implies $\Delta_-(p_1) < \Delta_-(p_2)$ for all $p_1, p_2$, and vice versa. For $k > 1$, reflection effect is not always observed. For example, both $\Delta(0.5)$ and $\Delta_-(0.5)$ are negative and $\Delta(1) = \Delta_-(1) = 0$. So, we have $\Delta(1) > \Delta(0.5)$ and $\Delta_-(1) > \Delta_-(0.5)$, violating the reflection effect. In general, the reflection effect can be observed for extreme values of $p_1$ and $p_2$, while it becomes ambiguous when they take intermediate values.

## Appendix 2: Reproducing the results for problem 10 in Kahneman's 1979 paper

Here we are showing that our sequential branching mechanism implies that for problem 10 in Kahneman's 1979 paper, consistent to the experiment observations and what PT proposed, it is effectively equivalent to the case when the 1st stage of the problem is ignored. The branching scheme of a general version of problem 10 is shown in Figure S1.

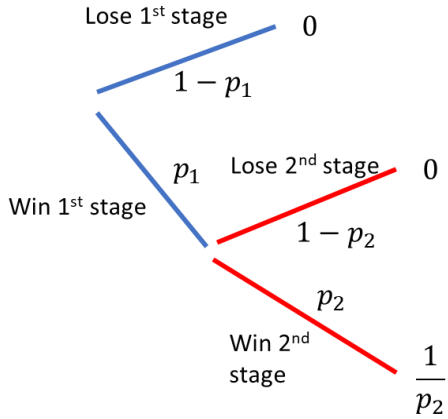

Figure S1: The branching scheme for a general version of problem 10 in Kahneman & Tversky (1979).

The surprise value $\Delta_{two\ stage}$, as a function $p_2$, is given by:

$$\Delta_{two\ stage}(p_2) = (1-p_1)\delta(0-p_1) + p_1\delta(1-p_1) + p_1\left(p_2\delta\left(\frac{1}{p_2} - 1\right) + (1-p_2)\delta(0-1)\right)$$

$$= f(1-p_1) - (1-p_1)kf(p_1) + p_1\left(p_2 f\left(\frac{1}{p_2} - 1\right) - (1-p_2)kf(1)\right) \qquad (A3)$$

Note that the term $f(1 - p_1) - k(1 - p_1)f(p_1)$ is independent of $p_2$. This is intuitive because events involving the same branch and same expected value at the intermediate states should generate the same anticipated surprise.

Let us consider two options. One with $p_2 = r$, the other with $p_2 = s$. The difference in the surprise value $D_{two\ stage} = \Delta_{two\ stage}(r) - \Delta_{two\ stage}(s)$ is given by:

$$D_{two\ stage} = p_1 \left( rf\left(\frac{1}{r} - 1\right) - sf\left(\frac{1}{s} - 1\right) - k(r - s)f(1) \right) \tag{A4}$$

In Kahneman's 1979 paper, it was suggested that the problem is equivalent to the case when the common blue branch is removed. When the common branch is removed, the problem reverts to the one that we discussed in Section 1. The surprise $\Delta_{one\ stage}(p_2)$ is given by eq. (A1) in Appendix 1. The difference in the surprise value for the same two options $D_{one\ stage} = \Delta_{one\ stage}(r) - \Delta_{one\ stage}(s)$ is given by:

$$D_{one\ stage} = rf\left(\frac{1}{r} - 1\right) - sf\left(\frac{1}{s} - 1\right) - k(r - s)f(1) \tag{A5}$$

The option preference is determined by the sign of $D$. Here, since $D_{two\ stage}$ only differs from $D_{one\ stage}$ by a multiplicative factor $p_1$, they always have the same sign. Thus, the model predicts the same preference no matters the choice of $r$ and $s$, meaning that the two-stage problem and the single-stage problem (obtained by removing the common branch) are practically equivalent in the anticipated surprise model.

## Appendix 3: Blackjack gambling

### The rule of Blackjack

Here we are only describing the rules of the blackjack that are relevant to this work. Also please note that there are no official rules for blackjack as casinos are free to introduce their own house rules. However, the one we describe here is one of the best known, widely used in casinos and work in gambling analysis (Shackleford, 2019a).

Blackjack uses standard 52-card decks. In a casino setting where many games are played, multiple decks of cards are used and played cards are introduced back into the deck often. This is to minimize the change of the winning odds as a result of changes in the composition of the deck as cards are exhausted. The goal of the player is to obtain a hand of cards with a higher value than that of the dealer. The value of the hand depends on a point system. Points are calculated by summing up the numbers on all cards in the hand. Face cards (i.e. Js, Qs and Ks) stands for 10 points. For the Aces, they can stand for either 1 point or 11 points, chosen in order to maximize the value of the hand. The value of a hand in descending order is as follows: an Ace and a single card with 10 points (also known as blackjack (abbrev. BJ)), 21 points (but not BJ), 20 points, 19 points, 18 points, 17 points, 4-16 points, more than 21 points (known as busted hand) for dealer, busted hand for player.

In the beginning of a round, the player is given two cards and the dealer is given one card faced up and one card faced down (the player cannot read the card faced down). The player plays before the dealer, except when the faced-up card of the dealer is an Ace, in which case we will cover later. He can choose to take an extra card or not to. If he chooses to take an extra card, a card will be dealt to him, and the same option will be presented to him again until his hand become busted. If he chooses not to, the turn will be passed to the dealer. The dealer will keep taking extra cards until his hand has more than or equal to 17 points. After the dealer finished playing, the round ends, and whoever has a hand with a higher value wins. When the player wins, he receives twice the amount of his original bet, thus giving him a net win of the size of his bet. An exception is when he wins with a hand of blackjack, in which case he will in addition get an extra amount of 0.5 times of his bet. When the player loses, he receives nothing, thus giving him a net loss of the size of his bet. In the event where the hand of the player and the dealer has equal value, normally the player will get back his bet and thus winning or losing nothing. However, to simplify our analysis, in section 3.1.1 and 3.1.2.2, we assume that the player will instead throw a coin to decide if he wins or loses, giving him 50% chance of winning and 50% chance of losing an amount equal to the size of his bet.

If the face-up card of the dealer is an Ace, the player will be given an option to place a side bet of the size half his original bet, known as 'the insurance', to bet on whether the hand of the dealer is Blackjack. Before the player starts playing, the dealer will peek at the face-down card. If the dealer has blackjack, the player will get a net win of the amount 2 times of his side bet, which is equivalent to the size of his original bet. At this point, the original bet can also be resolved. If the dealer does not have blackjack, the player will lose his side bets and play resumes in order to resolve the original bet. The placement of the insurance side bet does not affect how the original bet is resolved.

### Analysis for Section 3.11 (16 vs 10 situation)

Here we are showing that $\Delta_{stand} \geq \Delta_{hit}$. First, we note that the size of $x$ has no role in affecting the rank of $\Delta_{bet}$ and $\Delta_{not\,bet}$ since we can make a transformation to remove $x$ during the comparison between these two quantities. To simplify our analysis, we set $x = 1$ such that
$\Delta_{stand}$ and $\Delta_{hit}$ is given by:

$$\Delta_{stand} = p_0 f(1 - p_0) - (1 - p_0) k f(p_0) \tag{A6a}$$

$$\Delta_{hit} = -\left(1 - \frac{p_0}{p_2}\right) k f(p_0) + \frac{p_0}{p_2}[f(p_2 - p_0) + p_2 f(1 - p_2) - (1 - p_2) k f(p_2)] \tag{A6b}$$

$$D = \Delta_{stand} - \Delta_{hit} = p_0 \left(\frac{1}{p_2} - 1\right) k(f(p_2) - f(p_0)) + p_0(f(1 - p_0) - f(1 - p_2)) - \frac{p_0}{p_2} f(p_2 - p_0) \tag{A7}$$

Since $f$ is convex, $f(p_2 - p_0) \leq f(p_2) - f(p_0)$, eq. (A7) can then be rewritten as

$$D \geq \frac{p_0}{p_2}(k - 1)(f(p_2) - f(p_0)) + p_0[(f(1 - p_0) + k f(p_0)) - (f(1 - p_2) + k f(p_2))] \tag{A8}$$

For $k = 1$, the 1st term in eq. (A8) vanishes. For the 2nd term, define $F(z) = f(1 - z) + f(z)$, $z \in [0,1]$.

Consider the conditions $0 \leq p \leq q \leq \frac{1}{2}$, using the properties of a convex function again, we have

$$f(1 - p) - f(1 - q) \geq (q - p)f'(1 - q)$$
$$f(q) - f(p) \leq (q - p)f'(q) \leq (q - p)f'(1 - q) \leq f(1 - p) - f(1 - q) \quad \text{(A9)}$$

Eq. (A9) means $F(q) > F(p)$ for $0 \leq p \leq q \leq \frac{1}{2}$. For the domain $\frac{1}{2} < z \leq 1$, we can make use of the fact that by definition, $F(z) = F(1 - z)$.

In a player's 16 vs dealer's 10 situation, $p_0 = 0.23$, $p_1 = \frac{5}{13}$, $p_2 = \frac{p_0}{p_1} = 0.6$ (Shackleford, 2019b). By above, $f(1 - p_0) + f(p_0) = F(0.23) > F(0.4) = F(0.6) = f(1 - p_2) + f(p_2)$, and thus $D \geq 0$.

For $k > 1$, we note that $\Delta\delta$ is an increasing function with $k$ since
$$\frac{dD}{dk} = p_0\left(\frac{1}{p_2} - 1\right)\left(f(p_2) - f(p_0)\right) > 0 \quad \text{(A10)}$$

Therefore $D \geq 0$ also holds for $k > 1$.

## Analysis for Section 3.1.2.2 (taking side bets when the player has a non-blackjack good hand)

Here we are showing that $\Delta_{bet} \geq \Delta_{no\ bet}$. We are assuming that both the options of taking and not taking side bets has the same expected value, this amounts to $p_1 = \frac{1}{3}$ (in reality $p_1$ is slightly different at $p_1 = \frac{4}{13}$). Denoting the initial expected value of the hand, i.e before the dealer peeks for BJ by $E_0$. $E_0$ can be expressed as

$$E_0 = (1 - p_1)\left[\frac{x}{2}p_2 - \frac{3x}{2}(1 - p_2)\right] = \frac{2x}{3}\left(2p_2 - \frac{3}{2}\right) = x(\tfrac{4}{3}p_2 - 1) \quad \text{(A11a)}$$

Like in the previous section, the size of $x$ has no role in affecting the rank of $\Delta_{bet}$ and $\Delta_{not\ bet}$ since we can make a transformation to remove $x$ during the comparison between these two quantities. To simplify our analysis, we set $x = 1$ such that

$$E_0 = \tfrac{4}{3}p_2 - 1 \quad \text{(A11b)}$$

Let $E_1$ be the expected value after the dealer peeks for BJ if the player takes the side bet.
$$E_1 = \frac{E_0}{(1 - p_1)} = \tfrac{3}{2}E_0 \quad \text{(A12)}$$

Let $E_2$ be the expected value after the dealer peeks for BJ if the player does not take the side bet.
$$E_2 = E_1 + \tfrac{1}{2} = \tfrac{3}{2}E_0 + \tfrac{1}{2} \quad \text{(A13)}$$

It is trivial that $E_2 \geq E_0$. However, whether it is $E_1 \geq E_0$ or $E_1 \leq E_0$ depends on the sign of $E_0$. Therefore, we divide the problem in the separate cases: $E_0 < 0$ and $E_0 \geq 0$.

*Case 1: $E_0 < 0$*

When $E_0 < 0$, $E_1 < E_0$. $\Delta_{bet}$ and $\Delta_{no\ bet}$ can be expressed as

$$\Delta_{bet} = p_1 g f(-E_0) + (1-p_1)\left[-kf\left(\frac{-E_0}{2}\right) + p_2 f\left(\frac{1}{2} - \frac{3E_0}{2}\right) - (1-p_2)kf\left(\frac{3}{2} + \frac{3E_0}{2}\right)\right] \quad \text{(A14a)}$$

$$\Delta_{no\ bet} = -p_1 k f(1+E_0) + (1-p_1)\left[f\left(\frac{1}{2} + \frac{E_0}{2}\right) + p_2 f\left(\frac{1}{2} - \frac{3E_0}{2}\right) - (1-p_2)k_1 f\left(\frac{3}{2} + \frac{3E_0}{2}\right)\right] \quad \text{(A14b)}$$

Substituting in $p_1 = \frac{1}{3}$, we have

$$D = \Delta_{bet} - \Delta_{no\ bet} = \frac{1}{3}\left[(f(-E_0) + kf(1+E_0)) - 2\left(kf\left(\frac{-E_0}{2}\right) + f\left(\frac{1}{2}(1+E_0)\right)\right)\right] \quad \text{(A15)}$$

For $k = 1$, using the properties of a convex function, we have

$$f(-E_0) \geq 2f\left(\frac{-E_0}{2}\right) \text{ and } f(1+E_0) \geq 2f\left(\frac{1}{2}(1+E_0)\right).$$

It follows that $D \geq 0$.

For $k > 1$, we again study the derivative of $\Delta \delta$ with respective to $k$

$$\frac{dD}{dk} = \frac{1}{3}\left[f(1+E_0) - 2f\left(\frac{-E_0}{2}\right)\right] \quad \text{(A16)}$$

If $E_0 \geq \frac{-1}{2}$, we have $f(1+E_0) \geq 2f\left(\frac{1}{2}(1+E_0)\right) \geq 2f\left(\frac{-E_0}{2}\right)$ such that $D$ increases with $k$ and remains positive at $k > 1$.

From eq. (A11b), $E_0 \geq \frac{-1}{2}$ when $p_2 \geq \frac{3}{8}$. Since a large $p_2$ means that there is a large probability of winning with the hand, which, in order words, means the hand is a good hand. The results suggest that players prefer to take side bets when they have good hands.

*Case 2: $E_0 \geq 0$*

Since in the previous section, we established that $E_0 \geq \frac{-1}{2}$ is always considered good hands, we have to show that when $E_0 > 0$, $\Delta \delta \geq 0$ unconditionally. Noting that when $E_0 \geq 0$, $E_1 \geq E_0$, and that from eq. (A11b) $E_0$ has an upper bound of $\frac{1}{3}$, $\Delta_{bet}$ and $\Delta_{no\ bet}$ is given by

$$\Delta_{bet} = -p_1 k f(E_0) + (1-p_1)\left[f\left(\frac{E_0}{2}\right) + p_2 f\left(\frac{1}{2} - \frac{3E_0}{2}\right) - (1-p_2)kf\left(\frac{3}{2} + \frac{3E_0}{2}\right)\right] \quad \text{(A17a)}$$

$$\Delta_{no\ bet} = -p_1 k f(1+E_0) + (1-p_1)\left[f\left(\frac{1}{2} + \frac{E_0}{2}\right) + p_2 f\left(\frac{1}{2} - \frac{3E_0}{2}\right) - (1-p_2)kf\left(\frac{3}{2} + \frac{3E_0}{2}\right)\right] \quad \text{(A17b)}$$

Substituting in $p_1 = \frac{1}{3}$, we have

$$D = \Delta_{bet} - \Delta_{no\ bet} = \frac{1}{3}\left[k(f(1+E_0) - f(E_0)) - 2\left(f\left(\frac{1}{2} + \frac{E_0}{2}\right) - f\left(\frac{E_0}{2}\right)\right)\right] \quad \text{(A18)}$$

For $k = 1$, again using the properties of a convex function, we have

$$f(1 + E_0) - f\left(\frac{1}{2} + \frac{E_0}{2}\right) > \left(\frac{1}{2} + \frac{E_0}{2}\right) f'\left(\frac{1}{2} + \frac{E_0}{2}\right)$$

$$f\left(\frac{1}{2} + \frac{E_0}{2}\right) - f\left(\frac{E_0}{2}\right) < \frac{1}{2} f'\left(\frac{1}{2} + \frac{E_0}{2}\right)$$

$$f(E_0) - f\left(\frac{E_0}{2}\right) < \left(\frac{E_0}{2}\right) f'(E_0) < \frac{E_0}{2} f'\left(1 + \frac{E_0}{2}\right)$$

This gives

$$f(1 + E_0) - f\left(\frac{1}{2} + \frac{E_0}{2}\right) > f\left(\frac{1}{2} + \frac{E_0}{2}\right) - f\left(\frac{E_0}{2}\right) + f(E_0) - f\left(\frac{E_0}{2}\right)$$

$$f(1 + E_0) - f(E_0) > 2\left(f\left(\frac{1}{2} + \frac{E_0}{2}\right) - 2f\left(\frac{E_0}{2}\right)\right)$$

It follows that $D \geq 0$.

From (A18), it is obvious that $D$ increases with $k$ such that $D \geq 0$ still holds when $k > 1$.

## Appendix 4: Ellsberg paradox

Here we are exploring the condition in which the unambiguous urn would be preferred over the ambiguous urn. We assume that a symmetric prior probability that satisfies $p(m) = p(1 - m)$ for $m \in \left\{0, \frac{1}{2n}, \frac{2}{2n}, \ldots, 1\right\}$. The surprise value for picking the unambiguous urn ($\Delta_1$) and the ambiguous urn ($\Delta_2$) is given by eqs. (6a) and (6b) in the main text. For convenience, we are repeating them here:

$$\Delta_1 = \frac{1}{2}\delta\left(\frac{1}{2}\right) + \frac{1}{2}\delta\left(-\frac{1}{2}\right) = \frac{1}{2}f\left(\frac{1}{2}\right) - \frac{k}{2}f\left(\frac{1}{2}\right)$$

$$= \sum_m p(m)\left[\frac{1-k}{2} f\left(\frac{1}{2}\right)\right]$$

$$= (1 - k)\sum_{m<1/2} p(m) f\left(\frac{1}{2}\right) + p\left(\frac{1}{2}\right)\frac{1-k}{2} f\left(\frac{1}{2}\right) \tag{A19a}$$

$$\Delta_2 = \sum_m p(m)\left[\delta\left(m - \frac{1}{2}\right) + m\delta(1 - m) + (1 - m)\delta(-m)\right]$$

$$= \sum_{m<1/2} p(m)\left[-kf\left(\frac{1}{2} - m\right) + mf(1 - m) - k(1 - m)f(m)\right] + p\left(\frac{1}{2}\right)\frac{1-k}{2} f\left(\frac{1}{2}\right)$$

$$+ \sum_{m>1/2} p(m)\left[f\left(m - \frac{1}{2}\right) + mf(1 - m) - k(1 - m)f(m)\right]$$

$$= \sum_{m<1/2} p(m)\left[-kf\left(\frac{1}{2} - m\right) + mf(1 - m) - k(1 - m)f(m)\right] + p\left(\frac{1}{2}\right)\frac{1-k}{2} f\left(\frac{1}{2}\right)$$

$$+ \sum_{m<1/2} p(1 - m)\left[f\left(\frac{1}{2} - m\right) + (1 - m)f(m) - kmf(1 - m)\right]$$

$$= (1 - k)\sum_{m<1/2} p(m)\left[f\left(\frac{1}{2} - m\right) + mf(1 - m) - (1 - m)f(m)\right] + p\left(\frac{1}{2}\right)\frac{1-k}{2} f\left(\frac{1}{2}\right)$$

(A19b)

Note that we used $\sum_{m>1/2} F(1 - m) = \sum_{m<1/2} F(m)$ for general function $F$.

Hence,

$$D = \Delta_2 - \Delta_1 = (1-k) \sum_{m<1/2} p(m) \left[ f\left(\frac{1}{2} - m\right) + mf(1-m) - (1-m)f(m) - f\left(\frac{1}{2}\right) \right]$$
(A20)

*Condition 1: f is not strongly convex*

Since $k \geq 1$, a sufficient condition for $D > 0$ is

$$f\left(\frac{1}{2} - m\right) + mf(1-m) + (1-m)f(m) > f\left(\frac{1}{2}\right) \; \forall 0 \leq m \leq \frac{1}{2}$$
(A21)

Define $F(m) = f\left(\frac{1}{2} - m\right) + mf(1-m) + (1-m)f(m)$. Note that

$$F(0) = F\left(\frac{1}{2}\right) = f\left(\frac{1}{2}\right)$$

$$F'(m) = -f'\left(\frac{1}{2} - m\right) - mf'(1-m) + f(1-m) + (1-m)f'(m) - f(m)$$

$$F'\left(\frac{1}{2}\right) = -f'(0) < 0$$

Based on these results, since $f$ (and hence $F$) is a continuous function, (A21) would be true if $F''(m) < 0 \; \forall 0 \leq m \leq \frac{1}{2}$

$$F''(m) = (1-m)f''(m) + mf''(1-m) + f''\left(\frac{1}{2} - m\right) - 2f'(1-m) - 2f'(m) \quad \text{(A22)}$$

It is not true that $F''(m) < 0$ at $0 \leq m \leq \frac{1}{2}$ for any convex function $f$. Here we would explore possible conditions where $F''(x) < 0$ is true. We note that

$$F''(m) = (1-m)f''(m) + mf''(1-m) + f''\left(\frac{1}{2} - m\right) - 2f'(1-m) - 2f'(m)$$

$$= (1-m)[f''(m) - 2f'(m)] + m[f''(1-m) - 2f'(1-m)] - \left[2mf'(m) + (2-2m)f'(1-m) - f''\left(\frac{1}{2} - m\right)\right]$$
(A23)

Note that if $f''(m) \leq \frac{3}{2}f'(m)$, when $0 \leq m < \frac{1}{4}$,

$$2mf'(m) + (2-2m)f'(1-m) > \frac{3}{2}f'(1-m) \geq \frac{3}{2}f'\left(\frac{1}{2} - m\right) \geq f''\left(\frac{1}{2} - m\right),$$

and when $\frac{1}{4} \leq m \leq \frac{1}{2}$,

$$2mf'(m) + (2-2m)f'(1-m) \geq 2f'(m) \geq 2f'\left(\frac{1}{2} - m\right) > f''\left(\frac{1}{2} - m\right).$$

Therefore, one can infer from eq. (A23) that

$$F''(m) < 0 \text{ if } f''(m) \leq \frac{3}{2}f'(m) \; \forall 0 \leq m \leq \frac{1}{2}.$$
(A24)

Please note that eq. (A24) is a sufficient condition but not a necessary condition for $D > 0$. As we have shown in Figure 5, there are functions $f$ which do not fulfil eq. (A24) but still allows $D > 0$.

Eq. (A24) suggests that if is not very convex then, $D > 0$ is guaranteed. On the other hand, $D > 0$ can also be achieved if $f$ is strongly convex, as we will show below.

*Condition 2: f is strongly convex*

We investigate what happens if $f$ is strongly convex. Starting from eq. (A20),

$$D = (k-1)\sum_{m<1/2} p(m)\left[f\left(\frac{1}{2}-m\right) + mf(1-m) + (1-m)f(m) - f\left(\frac{1}{2}\right)\right]$$

$$> (k-1)\sum_{m<1/2} p(m)\left[mf(1-m) - f\left(\frac{1}{2}\right)\right] \qquad (A25)$$

because $mf(1-m)$ would dominate the other neglected terms for strongly convex $f$.

Next, we make use of the fact that $f$ is convex. There are multiple ways of doing it, which will lead to different constraints on $f$ for ambiguity aversion. Here, we show an example.

Applying the Jensen's inequality, we have $\mathrm{E}^{(1)}[f(1-m)] \geq f(1 - \mathrm{E}^{(1)}[m])$,

where $\mathrm{E}^{(c)}[\cdot] \equiv \frac{\sum_{m<1/2} \cdot\, m^c p(m)}{\sum_{m<1/2} m^c p(m)}$. This leads to

$$D > (k-1)\left(\sum_{m<1/2} p(m)\right)\left[\mathrm{E}^{(0)}[m]f\left(1 - \mathrm{E}^{(1)}[m]\right) - f\left(\frac{1}{2}\right)\right]. \qquad (A26)$$

Note that $0 < \mathrm{E}^{(c)}[m] < \frac{1}{2}$ since $0 < m < 1/2$ for both $c = 0$ and 1. Hence, $D > 0$ is guaranteed if $f$ is sufficiently convex. More specifically,

$$\mathrm{E}^{(0)}[m]f\left(1 - \mathrm{E}^{(1)}[m]\right) > f\left(\frac{1}{2}\right) \qquad (A27)$$

For example, for $f(m) = m^r$ and $p(m) = \frac{1}{2n+1}$ for all $m$, one can easily show that $D > 0$ if

$$r > \frac{\log\frac{4n}{n-1}}{\log\left(\frac{4n+1}{3n}\right)}. \qquad (A28)$$

which implies a strongly convex function $f$ with $r > 4.82$ in the limit of large $n$. We would like to emphasize again that eqs. (A27) and (A28) are not necessary condition for $D > 0$. In fact, as we have shown in our numerical study (Figure 5), $D > 0$ can be achieved with a much less strongly convex function than eqs. (A27) and (A28) suggest.

# Appendix 5: Analysis on CPT and RT on examples in Section 3

## Blackjack side-bet taking for good hands (Section 3.1.2.2)

The options and their outcomes, arranged in rank order, in reorganized in Table A3.

| Reward for Option 1 (Side bets) | Reward for Option 2 (No side bets) | Probability |
|---|---|---|
| $-\frac{3}{2}$ | $-1$ | $\frac{2}{3} - p$ |
| $0$ | $0$ | $\frac{1}{3}$ |
| $\frac{1}{2}$ | $1$ | $p$ |

Table A3: The possible outcomes and their probability for taking and not taking side bets. $p$ is the probability of winning the hand, where $p = \frac{2}{3}p_2$ in the main text.

## RT

For RT, anticipated regret $R$ is given by $R = \sum_i p_i\, Q(u(y_i) - u(x_i))$, where $x_i$ is the outcome for the chosen option, $y_i$ is the outcome if the other option is chosen, $p_i$ is the probability of the outcomes, $u$ is a concave, odd function, $Q$ is a convex, odd function (Bleichrodt & Wakker, 2015). The option that minimizes $R$ is chosen. Because of the symmetry in $Q$, only terms where $x_i > y_i$ need to be considered.

The regret for taking the side bets is given by

$$R_{side\ bets} = \frac{2}{3}p_2 Q\left(u(1) - u\left(\tfrac{1}{2}\right)\right) + \frac{2}{3}(1 - p_2) Q\left(u\left(\tfrac{3}{2}\right) - u(1)\right) \tag{A29}$$

Again, we have set $x = 1$ as in Appendix 3.

Since $u$ is concave, $u(1) > \frac{1}{2}\left(u\left(\tfrac{1}{2}\right) + u\left(\tfrac{3}{2}\right)\right)$, and hence $u(1) - u\left(\tfrac{1}{2}\right) > u\left(\tfrac{3}{2}\right) - u(1)$ such that $R_{side\ bets}$ is increasing with $p_2$. This means that RT predicts that players have less tendency to take side bets when they have good hands than when they have bad hands, which is at odds with experimental observations.

## CPT

The prospect $V$ for taking and not taking side bets is given by

$$V_{side\ bets} = w\left(\tfrac{2}{3} - p\right)v\left(-\tfrac{3}{2}\right) + \left(w(1 - p) - w\left(\tfrac{2}{3} - p\right)\right)v(0) + \left(w(1) - w(1 - p)\right)v\left(\tfrac{1}{2}\right) \tag{A30a}$$

$$V_{no\ side\ bets} = w(1 - p)v(-1) + +\left(w(1) - w(1 - p)\right)v(1), \tag{A30b}$$

where $v(x) = x^r$ with $r < 1$, $v(-x) = -kv(x)$ with $k > 0$, for all $x \geq 0$, and $w(p) = \frac{p^\gamma}{(p^\gamma + (1-p)^\gamma)^{\frac{1}{\gamma}}}$ with $\gamma < 1$. Please note that we have made the simplifying assumption that the the concavity of $v$ and the value of $\gamma$ is the same for the gain and loss domain. This simplification would not lead to qualitative difference in the analysis (results not shown). For details, please refer to Tversky & Kahneman (1992).

The difference in the prospect between the options $D_{CPT}$ is given by

$$D_{CPT} = V_{side\ bets} - V_{no\ side\ bets} = k\left(w(1-p)v(1) - w\left(\frac{2}{3}-p\right)v\left(-\frac{3}{2}\right)\right) - $$
$$\left(w(1) - w(1-p)\right)\left(v(1) - v\left(\frac{1}{2}\right)\right) \tag{A31}$$

We have plotted $D_{CPT}$ against $p$ for a variety of parameters $\gamma$, $k$ and $r$, as shown in Figure S2.

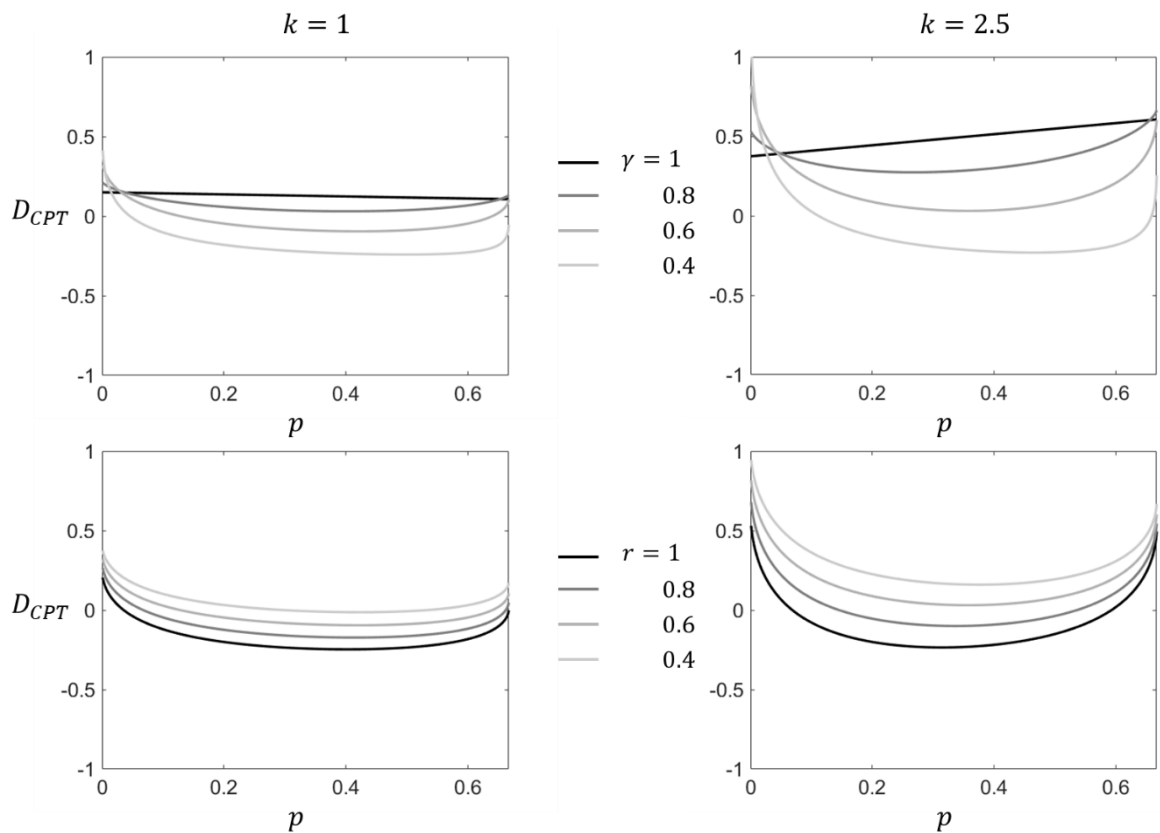

Figure S2: Differences in prospect between taking and not taking side bets ($D_{CPT}$) at different $p$ for different choices of parameters. Positive $D_{CPT}$ implies preference for taking side bets while large $p$ implies good hands. Top: $r$ is fixed at 0.6 while the lines in the plots correspond to different values of $\gamma$. Bottom: Top: $\gamma$ is fixed at 0.6 while the lines in the plots correspond to different values of $r$. Left: $k = 1$. Right: $k = 2.5$. Note that no combination of parameters is consistent with the experimental observation that taking side bets is preferred when the player has a good hand.

It can be observed that both the concavity of the value function $r$ and probability weighting $\gamma$ raised $D$ at small $p$. The latter also suppressed $D$ at large $p$. Overall, this is again at odds with experimental results where $D$ should be increasing with $p$ and should not be positive at all $p$.

As a comparison, the differences in anticipated surprise predicted by our model $D$ (See eq. (A15)) using realistic parameters we suggested at Section 3.3 in the main text are shown in Figure S3. Consistent with our analysis at Appendix 3 and the experimental observations, $D$ is increasing with $p$ and is positive at large $p$.

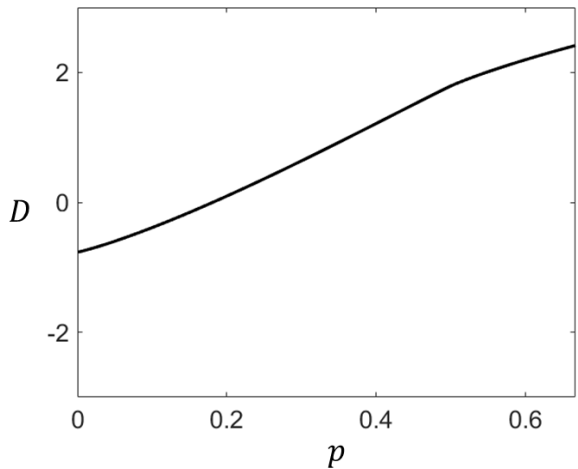

Figure S3: Differences in anticipated surprise based on the model in this work between taking and not taking side bets ($D$) at different $p$. Positive $D$ implies preference for taking side bets. Our model prediction is consistent with the experimental observation that taking side bets is preferred when the player has a good hand. We set $f(x) = x^r$, where $r = 1.5$, and $k = 2.5$.

## Allais paradox (Section 3.3)

For convenience, we state the problems in the original version of Allais paradox (i.e. Table 2) here again in Table A4.

(a) Problem 1

|  | Option 1 |  | Option 2 * |  |
|---|---|---|---|---|
|  | Reward | Probability | Reward | Probability |
| Outcome 1 | 0 | 0.89 | 0 | 0.9 |
| Outcome 2 | 1 | 0.11 | 5 | 0.1 |

(b) Problem 2

|  | Option 1 * |  | Option 2 |  |
|---|---|---|---|---|
|  | Reward | Probability | Reward | Probability |
| Outcome 1 | 1 | 1 | 0 | 0.01 |
| Outcome 2 |  |  | 1 | 0.89 |
| Outcome 3 |  |  | 5 | 0.1 |

Table A4: The two decision-making problems in the Allais paradox. The asterisk depicts the option preferred by most people.

## RT

For Problem 1, the regret for the options is given by:

$$R_{Option\ 1} = 0.089 Q(u(5)) + 0.011 Q(u(5) - u(1)) \tag{A32a}$$

$$R_{Option\ 2} = 0.099 Q(u(1)) \tag{A32b}$$

Option 2 is chosen implies that

$$0.089 Q(u(5)) + 0.011 Q(u(5) - u(1)) > 0.099 Q(u(1)) \tag{A33}$$

For Problem 2, the regret for the options is given by:

$$R_{Option\ 1} = 0.1 Q(u(5) - u(1)) \tag{A34a}$$

$$R_{Option\ 2} = 0.01 Q(u(1)) \tag{A34b}$$

Option 1 is chosen implies that $Q(u(1)) > 10 Q(u(5) - u(1))$ \hfill (A35)

If $u(5) < 2u(1)$, $u(5) - u(1) < u(1)$. eq. (A35) can be fulfilled by choosing a sufficiently convex function $Q$. To fulfil eq. (A33), a sufficient condition is $\frac{Q(u(5))}{Q(u(1))} > \frac{0.099}{0.089}$, which also requires $Q$ to be sufficiently convex.

These results show that RT is not theoretically inconsistent with Allais Paradox.

## CPT

For Problem 1, the prospect for the options is given by:

$$V_{Option\ 1} = w(0.9) v(0) + (w(1) - w(0.9)) v(5) \tag{A36a}$$

$$V_{Option\ 2} = w(0.89) v(0) + (w(1) - w(0.89)) v(1) \tag{A36b}$$

Option 2 is chosen implies that $v(5) > \frac{1 - w(0.89)}{1 - w(0.9)} v(1)$ \hfill (A37)

For Problem 2, the regret for the options is given by:

$$R_{Option\ 1} = w(0.01) v(0) + (w(0.9) - w(0.01)) v(1) + (1 - w(0.9)) v(5) \tag{A38a}$$

$$R_{Option\ 2} = v(1) \tag{A38b}$$

Option 1 is chosen implies that $v(5) < \frac{1 - w(0.9) + w(0.01)}{1 - w(0.9)} v(1)$ \hfill (A39)

One can easily see that eqs. (A37) and (A39) can be simultaneously fulfilled by imposing the condition $\frac{1 - w(0.89)}{1 - w(0.9)} < \frac{v(5)}{v(1)} < \frac{1 - w(0.9) + w(0.01)}{1 - w(0.9)}$, which, despite posing strict restriction on the parameters in $v$ and $w$, is not theoretically inconsistent.

# Reference (Appendices)

# Supplementary figures:

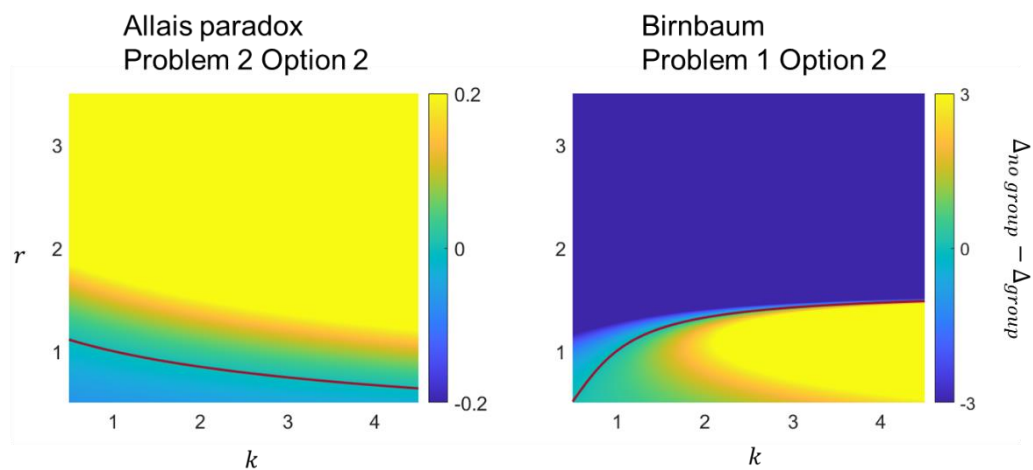

Figure S4: The difference in surprise values between the branching scheme when event grouping is present and when it is absence for Option 2 in the Allais paradox Problem 2 (left) and Option 2 in Problem 1 adopted from Birnbaum (2008) (right), which are discussed in the main text. The dark red line is the boundary where the options are equally preferred, i.e. $\Delta_{group} = \Delta_{no\ group}$. Consistent with our analysis, event grouping reduces the appeal for Option 2 in the Allais paradox for all $k, r > 1$. On the other hand, event grouping increases the appeal for Option 2 in the Birnbaum problem with a slightly more stringent condition that $r$ is not too close to 1 and $k$ is not too large. As with the rest of our numerical studies on these problems, we set $f(x) = x^r$.